\documentclass[fleqn,usenatbib]{rasti}

\usepackage{newtxtext,newtxmath}

\usepackage[T1]{fontenc}

\DeclareRobustCommand{\VAN}[3]{#2}
\let\VANthebibliography\thebibliography
\def\thebibliography{\DeclareRobustCommand{\VAN}[3]{##3}\VANthebibliography}

\usepackage{graphicx}	
\usepackage{amsmath}	

\usepackage{float} 
\usepackage{comment}
\usepackage{hyperref}

\title[CO cooling rates for AGB outflows]{A rotational line CO cooling rate prescription for AGB outflows}

\author[T. Ceulemans]{
T. Ceulemans,$^{1}$\thanks{E-mail: thomas.ceulemans@kuleuven.be}
F. De Ceuster,$^{2}$
O. Vermeulen$^{1}$
L. Decin$^{1}$
\\
$^{1}$Institute of Astronomy, Celestijnenlaan 200D, 3001 Leuven, Belgium\\
$^{2}$Leuven Gravity Institute, Celestijnenlaan 200D, 3001 Leuven, Belgium
}

\date{Accepted XXX. Received YYY; in original form ZZZ}

\pubyear{2025}

\begin{document}
\label{firstpage}
\pagerange{\pageref{firstpage}--\pageref{lastpage}}
\maketitle

\begin{abstract}
Asymptotic Giant Branch (AGB) stars significantly contribute to the chemical composition of the universe. 
In their outflows, complex chemistry takes place, which critically depends on the local temperature. 
Therefore, if we want to accurately model the AGB environment, we need accurate cooling rates. 
The CO molecule is abundant in AGB outflows, and has a dipole moment, which enables it to cool through emission from its rotational transitions. 
We therefore expect it to significantly contribute to cooling in this environment, even at low temperatures ($10$ K $\leqslant T\leqslant 3000$ K). 
Currently, CO cooling rates are available for ISM-like conditions, which encompasses a different parameter regime, with generally lower densities and velocity gradients, compared to AGB winds. 
Therefore, these ISM cooling rates might not be applicable to the AGB regime. In this paper, we compute CO cooling rates for hydrodynamics simulations of AGB outflows. 
To evaluate the net cooling rate, we calculate the energy level distribution of CO self-consistently, using the non Local Thermodynamical Equilibrium (NLTE) line radiative transfer code \textsc{Magritte}.
We verify whether already existing CO cooling rate prescriptions for the interstellar medium (ISM) are applicable for this regime. 
We noticed minor differences between these prescriptions and our calculated cooling rates in general. However, when used far outside their originally intended parameter regimes, significant differences occur. 
Therefore, we propose a new CO cooling rate prescription for the AGB environment and we study how the computed cooling rate varies depending on input parameters. 
	
\end{abstract}


\section{Introduction}\label{section: Introduction}

Asymptotic Giant Branch (AGB) stars are low-to-intermediate mass stars at the end of their lives. It is estimated that $\sim 85\%$ of dust and $\sim 35\%$ of gas in the interstellar medium originate from their outflows \citep{tielens_physics_2005}. In order to numerically evaluate the contribution of AGB outflows on the chemical evolution of the universe, we need to apply non-equilibrium chemistry to the stellar wind \citep[see e.g.][]{maes_sensitivity_2023}. An important input for the chemical network is the temperature, which is influenced by hydrodynamics, chemical cooling, and also by radiation transport.
In this paper, we focus on the cooling impact of CO rotational transition lines, as CO is the second most abundant molecule in the AGB outflows, aside from H$_2$. In contrast to H$_2$, CO has a net dipole moment, and thus easily excitable rotational energy transitions. Therefore CO can enable radiative cooling even at lower temperatures.

Radiative cooling forms an important part of the thermal balance of astronomical objects, starting from the interstellar medium \citep{neufeld_radiative_1993} to supernovae \citep{mcleod_carbon_2024}. However, no cooling prescriptions have yet focused on the AGB environment. The environment most similar to AGB outflows is the Interstellar Medium (ISM), for which cooling rates prescription have been computed in various publications, see e.g. \cite{neufeld_radiative_1993}, \cite{neufeld_thermal_1995}, \cite{omukai_low-metallicity_2010} and \cite{whitworth_simple_2018}.
In previous papers, simplifying assumptions have been made about the geometry of such objects, including spherical symmetry. However, for the AGB regime, we cannot assume such simplifying geometry, as significant asymmetries have been observed in the outflows \citep{decin_evolution_2020}.
Furthermore, the physical conditions in the ISM are different from the conditions in the AGB outflow, which can limit the applicability of the previously mentioned cooling rate prescriptions.

In this paper, we will compute CO cooling rates based on hydrodynamics models of AGB winds simulated with \textsc{Phantom} \citep{price_phantom_2018}. For this, we use \textsc{Magritte} \citep[see e.g.][]{deceuster_magritte_2019, deceuster_magritte_2020, de_ceuster_3d_2022, ceulemans_magritte_2024} in order to calculate the non Local Thermodynamical Equilibrium (NLTE) radiation field, which determines the cooling. Afterwards, we will compare the calculated cooling rates with pre-existing prescriptions \citep{neufeld_radiative_1993, neufeld_thermal_1995, omukai_low-metallicity_2010, whitworth_simple_2018}. However, we find that these cooling prescriptions are not fully suited to the AGB environment, when used outside their intended parameter regime. Therefore, we propose a new fit in this paper.

This paper is structured as follows: a brief introduction on how to calculate CO cooling rates is given in Section \ref{section: theory}. We introduce the literature cooling rate prescriptions relevant for this paper in Section \ref{section: existing cooling prescriptions}. We explain the model setup and calculate cooling rates in Section \ref{section: CO cooling rates for AGB outflows}. Afterwards, we discuss the impact of perturbations on the model input to the computed cooling rate in Section \ref{section: model perturbations}. Finally, we discuss the limitations of our cooling fits in Section \ref{section: limitations}.

\section{Computing the cooling rate}\label{section: theory}
In order to compute the net radiative cooling rate $\Lambda [\text{W m}^{-3}]$, we need to create radiative transfer models. For this, we use \textsc{Magritte} \citep{de_ceuster_3d_2022}. As input for our models, we use hydrodynamics models of AGB winds, which contain temperatures, densities, and a velocity field. We interpret the density as H$_2$ density and impose a constant CO/H$_2$ ratio, which is valid in the outflow as long as no significant photo-dissociation occurs. We then use this \textsc{Magritte} model to compute NLTE line radiative transfer. 

Our workflow is as follow: our initial guess for the energy level distribution of the species CO is given by the assumption of local thermodynamical equilibrium (LTE). We iteratively compute the radiation field $I(\hat{\boldsymbol{n}}, \boldsymbol{x}, \nu) [\text{W sr}^{-1} \text{m}^{-2} \text{Hz}^{-1}]$ based on the level populations $n_i [\text{m}^{-3}]$, and the level populations using statistical equilibrium. To speed up this iterative process, we use Accelerated Lambda Iteration (ALI) \citep[see e.g.][]{rybicki_accelerated_1991} and adaptive Ng-acceleration \citep[as described in][]{ceulemans_magritte_2024}. At the end of this iterative process, we have obtained self-consistent values for the radiation field $I$ and the level population $n_i$. Then, in order to calculate the net radiative cooling rate $\Lambda$, we have two options.
We subtract the emitted intensity from the absorbed intensity in order to obtain the net cooling rate $\Lambda$, as done in \cite{carlsson_approximations_2012},
\begin{equation}\label{eq: carlsson cooling}
\Lambda = \int \chi(\nu)(S(\nu)-I(\hat{\boldsymbol{n}}, \boldsymbol{x}, \nu)) \text{d}\nu \text{d}\Omega.
\end{equation}
In which $\chi[\text{m}^{-1}]$ is the opacity, $S [\text{W sr}^{-1} \text{m}^{-2} \text{Hz}^{-1}]$ is the source function and $I$ is the intensity. 
In high optical depth regimes, the intensity $I$ is almost equal to the source function $S$, but not exactly. Therefore Eq. \ref{eq: carlsson cooling} can suffer from numerical cancellation errors. In line radiative transfer, an equivalent prescription can be found when assuming statistical equilibrium. In this case, the equation from \cite{sahai_new_1990} is valid:

\begin{equation}
\Lambda = \sum_{\text{transitions from level i to j}} (n_jC_{ji}-n_iC_{ij})h\nu_{ij}\label{eq: sahai cooling rates}
\end{equation}
in which $n_i$ are the number densities at the different energy levels and $C_{ij} [\text{s}^{-1}]$ are the collisional rates. In this paper, we calculate the cooling rate $\Lambda$ using the latter equation.

\section{Existing cooling prescriptions}\label{section: existing cooling prescriptions}

In the literature, several CO molecular line cooling prescriptions already exist. However, they have been mainly focused on ISM-like conditions. As the environment around AGB stars is significantly different from the ISM conditions, with higher densities\footnote{Molecular clouds can have similar densities, but have lower temperatures when compared to the denser regions of the AGB outflow.} and velocity gradients, we first check whether the current prescriptions are valid in this environment. We consider the prescriptions in \cite{neufeld_radiative_1993} \cite{neufeld_thermal_1995}, \cite{omukai_low-metallicity_2010} and \cite{whitworth_simple_2018} for this paper. These prescriptions mostly use the following parameters: the $\text{H}_2$ and CO number densities $n_{\text{H}_2}, n_{\text{CO}} [\text{m}^{-3}]$, the local gas temperature $T [\text{K}]$ and the velocity divergence $\left|\nabla\cdot \boldsymbol{v}\right| [\text{s}^{-1}]$, which is related to the Sobolev optical depth $\tau_{\text{Sobolev}} [.]$ \citep{sobolev_moving_1960},
\begin{align}\label{eq: sobolev optical depth}
\tau_{\text{Sobolev}} = \chi\frac{c}{\left|\nabla\cdot \boldsymbol{v}\right|},
\end{align}
in which $c$ is the speed of light and $\chi$ is the frequency integrated opacity.
The densities directly affect the cooling rates (see Eq. \ref{eq: sahai cooling rates}), as the CO density affects the level populations $n_i$ of CO, and the H$_2$ density is used for the calculation of the collisional rates. The temperature determines the relative distribution of the level populations in LTE, and still impacts them in NLTE. The velocity divergence $\left|\nabla\cdot\boldsymbol{v}\right|$ limits how much radiation can escape from the molecular emission lines and therefore imposes an upper bound on the cooling rate \citep{neufeld_radiative_1993}.

In the following subsections, we briefly explain the literature cooling rate prescriptions explored in this paper. We note, however, that we will make different assumptions during our computations, calculate the cooling in a different astrophysical regime, and use different data when computing the cooling rates. Therefore, any difference we observe with the net cooling rate we compute does not invalidate their applicability to their intended astrophysical regime.

\subsection{Neufeld et al. 1993 \& 1995}


In \cite{neufeld_radiative_1993}, cooling rates have been calculated for H$_2$O, H$_2$ and CO using an escape probability method to compute the level populations. 
They assume the velocity field to have a large gradient and to be either monotonically increasing or decreasing, such that the Sobolev approximation is valid \citep[][]{sobolev_moving_1960}. In this way, NLTE radiative transfer can be computed using only local information.
Their rotational line cooling for CO is based on the first 75 rotational transitions, with energy levels taken from \textsc{HITRAN} database \citep{rothman_hitran_1987}, and uses the collision rate coefficients from \cite{viscuso_co_1988}.
Their rotational cooling prescription is valid between temperatures of $100$ K until $2000$ K and values of $\tilde{N}(\text{CO})$ between $10^{15} \text{s m}^{-3}$ en $10^{20} \text{s m}^{-3}$ \citep{neufeld_excitation_1991}, which they define as
\begin{align}
\tilde{N}(\text{CO})=\frac{n_{\text{CO}}}{\left|\nabla\cdot \boldsymbol{v}\right|},
\end{align}
in which $n_{\text{CO}}$ is the CO number density.
Their cooling prescription is given by
\begin{align}
\Lambda = L_{\text{N93}}n_{\text{H}_2}n_{\text{CO}}, 
\end{align}
in which the cooling rate coefficient $L_{\text{N93}}[\text{W m}^{3}]$ is given by
\begin{align}
\frac{1}{L_{\text{N93}}} = \frac{1}{L_0} + \frac{n_{\text{H}_2}}{\mathcal{L}_{\text{LTE}}} + \frac{1}{L_0}\left[\frac{n_{\text{H}_2}}{n_{1/2}}\right]^\alpha\left(1-\frac{n_{1/2}L_0}{\mathcal{L}_{\text{LTE}}}\right)
\end{align}
and the coefficients $L_0$, $\mathcal{L}_{\text{LTE}}$, $n_{1/2}$ and $\alpha$ are tabulated in \cite{neufeld_radiative_1993}.
In \cite{neufeld_thermal_1995}, they use the same methodology as \cite{neufeld_radiative_1993} to provide cooling rates for the following species: H$_2$, H$_2$O, CO, O$_2$, HCl, C, and O. Their cooling prescription for CO is valid for temperatures between $10$ K and $100$ K  and values of $\tilde{N}(\text{CO})$ between $10^{15.5} \text{s m}^{-3}$ en $10^{20} \text{s m}^{-3}$. We combined both prescriptions into one, respectively using \cite{neufeld_radiative_1993} or \cite{neufeld_thermal_1995} whenever the local temperature $T$ is higher than $100$ K or not.

\subsection{Omukai et al. 2010}
\cite{omukai_low-metallicity_2010} models the ISM during collapse into a protostar. To calculate the CO cooling, they also use an escape probability method to compute the energy level distribution \citep[see][]{omukai_protostellar_2000} in NLTE, and adopt the CO molecular data from the \textsc{LAMDA} database \citep{schoier_atomic_2005}, but do not specify which version of the data they use.
In contrast to \cite{neufeld_radiative_1993}, they do not use an approximation for the escape probability, but evaluate it using the column density of their spherically symmetric model.

Their CO cooling rate prescription is based on \cite{neufeld_radiative_1993}, with the difference that the parameter range for the temperature has a lower bound of $10$ K and their definition $\tilde{N}$ has been changed to $\tilde{N}=N/\sqrt{2k_BT/m_\text{CO}}$, in which $N [\text{m}^{-2}]$ is the column density, $k_B [\text{J K}^{-1}]$ is Boltzmann’s constant and $m_\text{CO} [\text{kg}]$ is the CO particle mass. 
This quantity can be derived from the optical depth $\tau$ (ignoring constant factors), when assuming no velocity field to be present,
\begin{align}
\tau = \int \chi_{ij}\phi_{ij} dx \sim \frac{N}{\delta\nu_{ij}} \sim \frac{N}{\sqrt{2k_BT/m_{\text{CO}}}}.
\end{align}
in which $\chi_{ij} [\text{Hz m}^{-1}]$ is the integrated line opacity, $\phi_{ij} [\text{Hz}^{-1}]$ is the line profile function and $\delta\nu_{ij} [\text{Hz}]$ is the line width.
However, \cite{omukai_low-metallicity_2010} do not mention why they changed the definition of $\tilde{N}$. For this paper, we evaluate their cooling fit using the definition of Neufeld for $\tilde{N}$ instead, given that significant velocity fields are present in our models.

\subsection{Whitworth \& Jaffa 2018}
In \cite{whitworth_simple_2018}, they aim to model the effect of CO cooling on the structure of molecular clouds in the ISM. As input for their cooling fit, they use the cooling rate results of \cite{goldsmith_molecular_1978}. In that paper, the authors use an escape probability method to determine the level populations $n_i$ and the net cooling rate $\Lambda$, and utilize a Sobolev approximation \citep{sobolev_moving_1960} to evaluate the optical depth $\tau$. This allows them to approximate NLTE radiative transfer locally.

In contrast to the previous cooling prescriptions, \cite{whitworth_simple_2018} deduce their cooling rate prescription from first principles in two separate regimes, defining a theoretically motivated fit function for both the low and high density regime. In the low density regime, they find 
\begin{align}
\Lambda_{\text{LO}} = \lambda_{\text{LO}} X_\text{CO} n_{\text{H}_2}^2 T^{3/2},
\end{align}
in which $\lambda_{\text{LO}}$ is the fit coefficient, $X_\text{CO} = n_\text{CO}/n_{\text{H}_2}$, $n_{\text{H}_2}$ is the H$_2$ density, while in high densities, they obtain
\begin{align}
\Lambda_{\text{HI}} = \lambda_{\text{HI}} T^4 \left|\nabla\cdot \mathbf{v}\right|,
\end{align}
in which $\lambda_{\text{HI}}$ is the fit coefficient.
Afterwards, they combine both formulae using
\begin{align}
\frac{\Lambda}{n_\text{CO}} = \left(\left(\frac{\Lambda_{\text{LO}}}{n_\text{CO}}\right)^{-1/\beta} + \left(\frac{\Lambda_{\text{HI}}}{n_\text{CO}}\right)^{-1/\beta}\right)^{-\beta}
\end{align}
in which $\beta$ is a fit parameter. Their fit prescription is valid for temperatures between $10$ and $60$ K and for values of $\frac{n_{\text{CO}}/n_{\text{H}_2}}{\left|\nabla\cdot \boldsymbol{v}\right|}$ between  $10^5$ and $10^9$ s.

All cooling rate prescriptions described in this section treat radiative transfer using the escape probability method, and therefore limit the geometry to which it can be applied. Given that we calculate radiative transfer on 3D AGB wind models, we use a different calculation method, and might therefore obtain slightly different results for the net cooling rate.

\section{CO Cooling rates for AGB outflows}\label{section: CO cooling rates for AGB outflows}
The goal of this paper is to obtain a CO cooling rate prescription for AGB binary simulations. We therefore took a variety of \textsc{Phantom} \citep{price_phantom_2018} AGB binary simulations with various circular orbit configurations and also a single star model for verification purposes. The \textsc{Phantom} smoothed particle hydrodynamics (SPH) models generate a wind by periodically launching particles from the star \citep[as explained in][]{siess_3d_2022} with initial velocity as given in Table \ref{table: phantom models}. All models use a mass loss rate of $1\cdot 10^{-7} M_\odot/$yr. The final binary model in this Table, v15a35, is taken from \cite{malfait_sph_2024}, where it was named v15m06.

\begin{table}
	\centering
	\caption{Model properties of the \textsc{Phantom} models used in this paper. $v_{\text{in}}$ denotes the initial velocity, $M_\star$ denotes the stellar mass, $M_{\text{com}}$ denotes the companion mass and $a$ the semi-major axis.}\label{table: phantom models}
	\begin{tabular}{|l | r | r | r | r|}
		\hline
		Name & $v_{\text{in}}$ [km/s]& $M_\star$ [$M_\odot$]& $M_{\text{com}}$ [$M_\odot$]& a [AU]\\
		\hline
		v05a06 & $5$ & $1.5$ & $1$ & $6$ \\
		v05a20 & $5$ & $1.5$ & $1$ & $20$ \\
		v20a06 & $20$ & $1.5$ & $1$ & $6$\\
		v20a20 & $20$ & $1.5$ & $1$ & $20$\\
		v15a35 & $15$ & $2$ & $0.6$ & $35$\\
		\hline
		v10\_single & $10$ & $1.5$ &\multicolumn{2}{c}{} \\
		\cline{1-3}
	\end{tabular}
\end{table}

\subsection{General model setup}\label{section: general model setup}
Starting from a \textsc{Phantom} SPH model, we calculate the NLTE CO level populations using Magritte \citep{ceulemans_magritte_2024, de_ceuster_3d_2022}, reducing the model size for increased computational efficiency, using the procedure detailed in \cite{ceulemans_magritte_2024}, with maximal recursion level $l=12$ and maximal variation $r_{\text{max}} = 0.2$. Compared to using a Sobolev approximation \citep{sobolev_moving_1960} for determining the NLTE level populations, as done in previous works (see Section \ref{section: existing cooling prescriptions}), the approach we use is more generally applicable to asymmetric 3D environments, given that it can also take into account non-local contributions to the intensity.
We utilize the $^{12}$CO line data from the \textsc{LAMDA} database\footnote{available at \url{https://home.strw.leidenuniv.nl/~moldata/}} \citep{schoier_atomic_2005}, in which the collision rates are taken from \cite{yang_rotational_2010}, extrapolated to $75$ levels as detailed in Appendix \ref{appendix: CO collisional rate extrapolation}. We only consider rotational transitions in this paper, as collisional data for the CO vibrational transitions is currently only available up to $300$ K \citep{yang_full-dimensional_2016}.
For each of the $37$ hydrodynamics snapshots from these AGB models, taken at different times, we generate different \textsc{Magritte} models, each with a different, but constant CO$/$H$_\text{2}$ abundance $\in \left\{1, 3, 5, 8, 10\right\}\cdot 10^{-4}$. This range spans the values used in the literature \citep[see e.g.][]{khouri_wind_2014, danilovich_detailed_2014} for the AGB envelope. We have limited our models to a radius of $500$ AU around the primary star, where we assume no significant CO dissociation to have taken place. In this way, a constant CO$/$H$_\text{2}$ abundance ratio should be a valid assumption. For most of this paper, we restrict ourselves to only using the binary models, because including this data would add too many data points in less interesting regions of the parameter space. We will however validate in Appendix \ref{appendix: applying full fit to single star} that our obtained fit of Section \ref{section: fit all} works on the single star data.

Our combined data set of AGB binary models contains $185$ \textsc{Magritte} models, having $18$ million data points in total, with each data point consisting of a vector containing the gas temperature $T$, H$_2$ density $n_{\text{H}_2}$, CO density $n_{\text{CO}}$, velocity gradient $\left|\nabla\cdot \boldsymbol{v}\right|$ and computed net cooling rate $\Lambda$.
The parameter ranges spanned by these models can be found in Table \ref{table: SPH parameter limits}. 
As inputs for the fit function, densities, temperatures and the velocity divergence $\left|\nabla\cdot \boldsymbol{v}\right|$ at each position are required, all of which can be extracted from the SPH model.

\begin{table}
	\centering
	\caption{Parameter limits for this paper}\label{table: SPH parameter limits}
	\begin{tabular}{|l|r|r|}
		\cline{2-3}
		\multicolumn{1}{c|}{}& Min & Max\\
		\hline
		Temperature [K]$\dagger$ & $14$ & $3000$\\
		H$_2$ density [m$^{-3}$]&$1.4\cdot 10^8$&$1.9\cdot 10^{15}$\\
		CO/H$_2$ ratio& $10^{-4}$& $10^{-3}$\\
		$n(\text{CO})/\left|\nabla \cdot v\right|$ [s m$^{-3}$]&$1.5\cdot 10^{13}$ &$4.2\cdot 10^{23}$\\
		\hline
	\end{tabular}
	
	\footnotesize
	$\dagger$ Limited by CO collisional data. SPH model temperatures can exceed $3000$K.
\end{table}

\subsection{Existing cooling rate prescriptions}\label{section: pre-existing cooling rate prescriptions tested}

\begin{figure}
	\includegraphics[width=\linewidth]{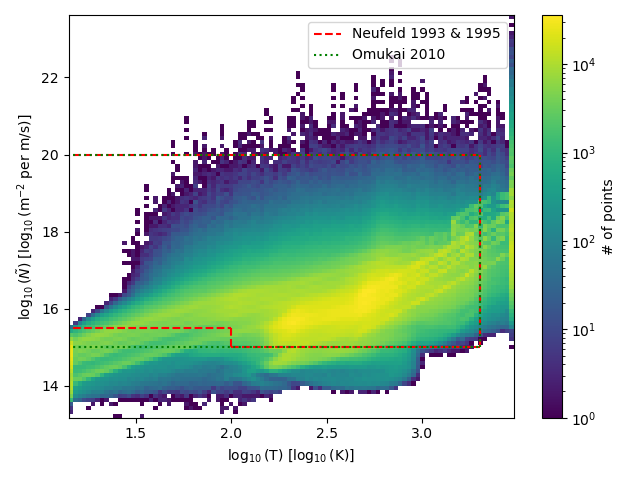}
	\caption{Distribution of our model points, compared to the parameter ranges in which the Neufeld and Omukai prescriptions are valid. A significant fraction of our data points lie outside the ranges covered by \protect\cite{neufeld_radiative_1993} and \protect\cite{omukai_low-metallicity_2010}.}
	\label{figure: parameter_bounds_neufeld_omukai}
\end{figure}

\begin{figure}
	\includegraphics[width=\linewidth]{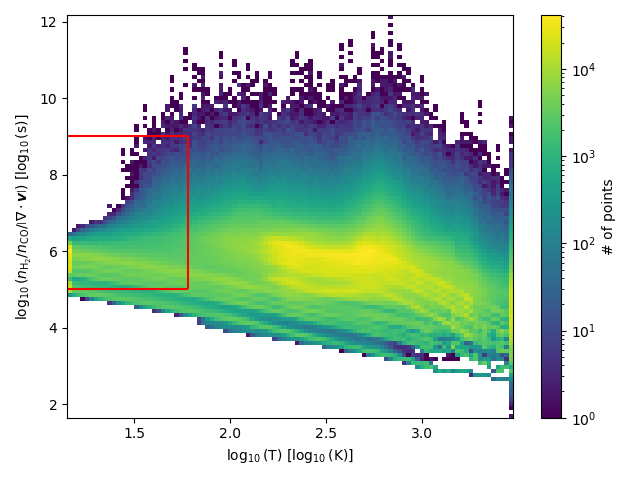}
	\caption{Distribution of our model points, compared to the parameter range (box) in which the Whitworth prescription is valid. Most of our data points lie outside the range covered by \protect\cite{whitworth_simple_2018}.}
	\label{figure: parameter_bounds_whitworth}
\end{figure}

\begin{figure}
	\includegraphics[width=\linewidth]{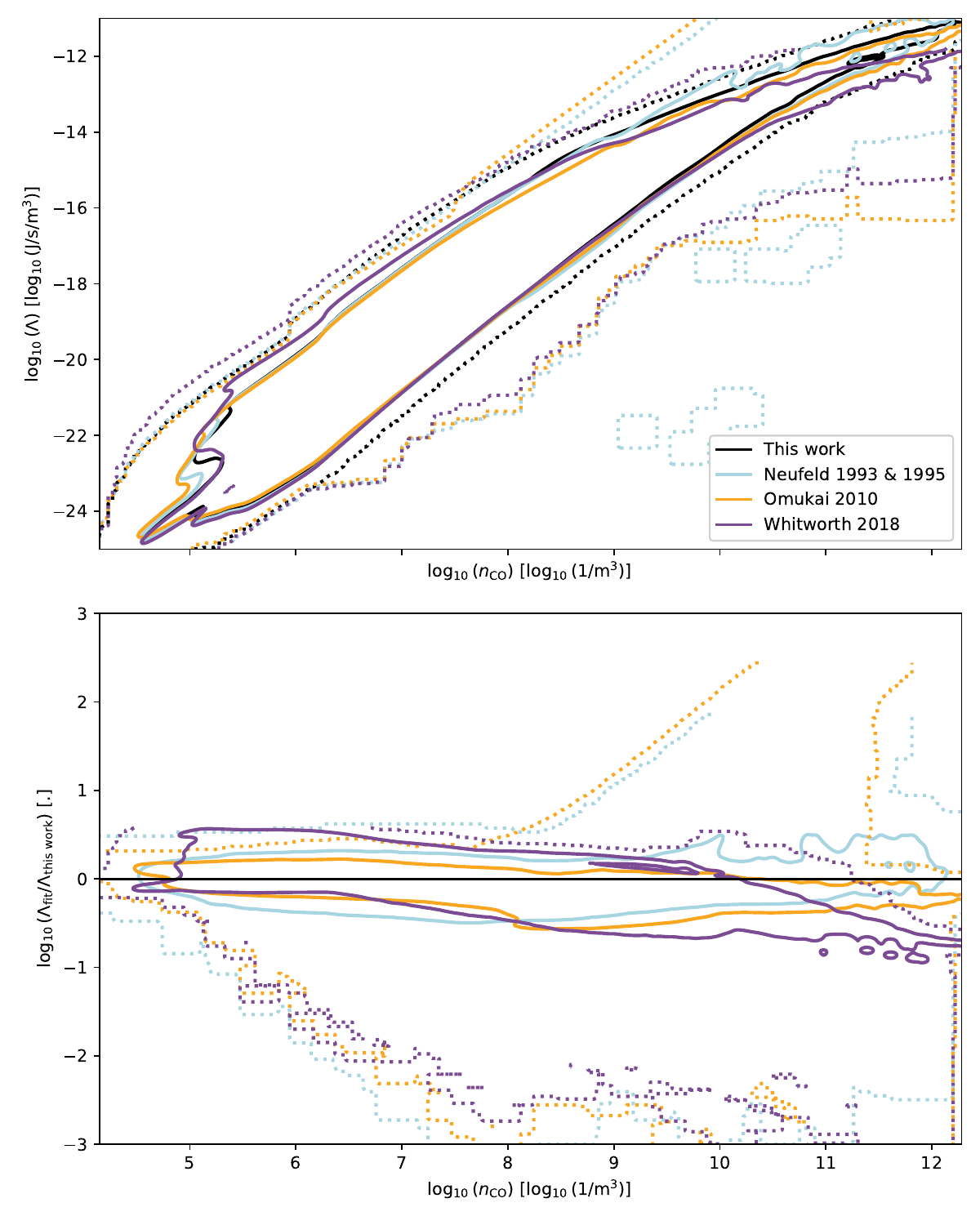}
	\caption{Contour plots of the cooling rate versus the CO density, smoothed using a Gaussian kernel. The solid line contains $99\%$ of the data points. The dotted line denotes the outer contour where data is present. Bottom: contour plot of the relative differences between the fits and our computed cooling rate.}
	\label{figure: contour plot co density}
\end{figure}

\begin{figure}
	\includegraphics[width=\linewidth]{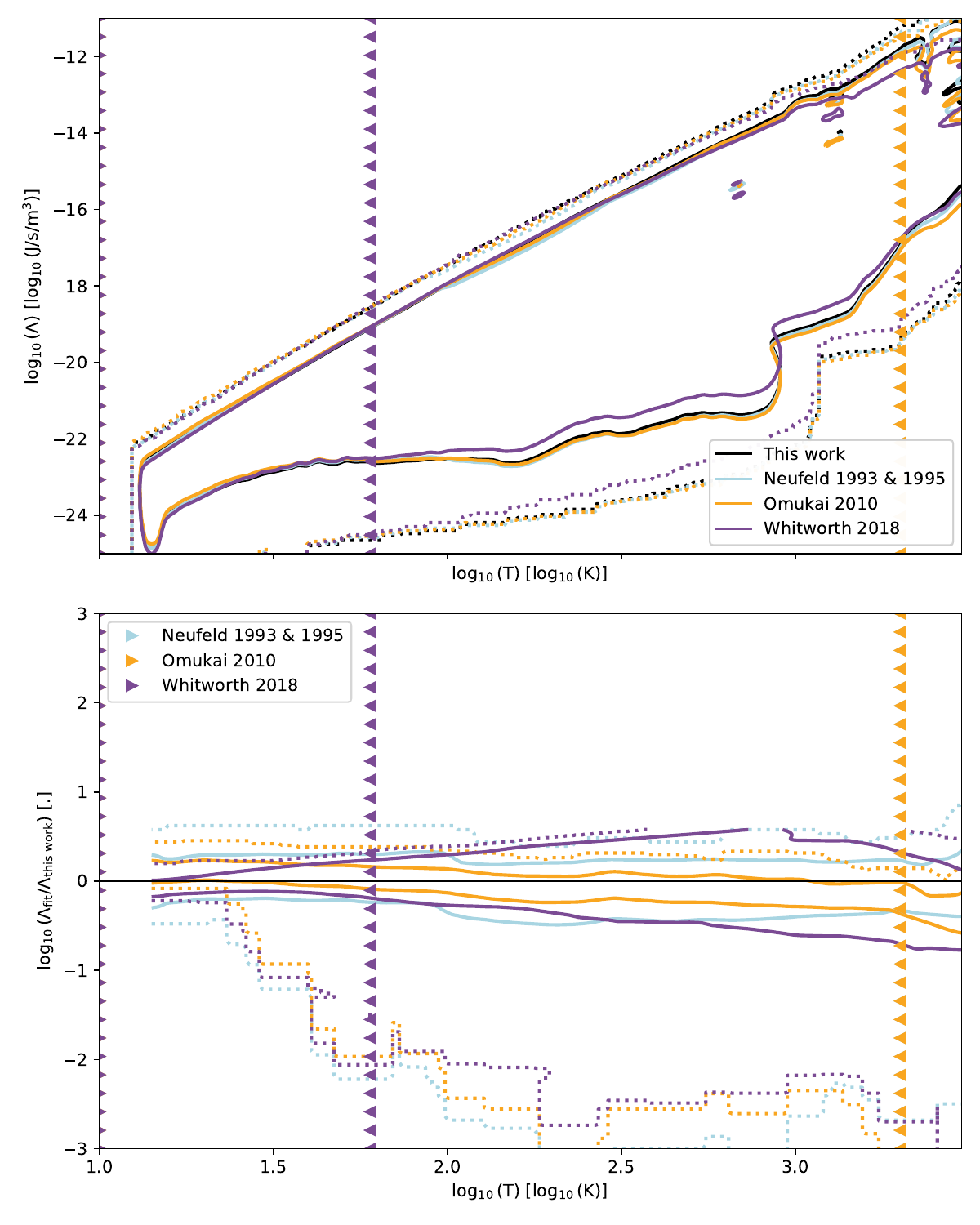}
	\caption{Contour plots of the cooling rate versus the temperature, smoothed using a Gaussian kernel. The solid line contains $99\%$ of the data points. The dotted line denotes the outer contour where data is present. The vertical arrow lines denote the temperature bounds for which the literature fit functions are defined. Bottom: contour plot of the relative differences between the fits and our computed cooling rate.}
	\label{figure: contour plot temperature}
\end{figure}

\begin{figure}
	\includegraphics[width=\linewidth]{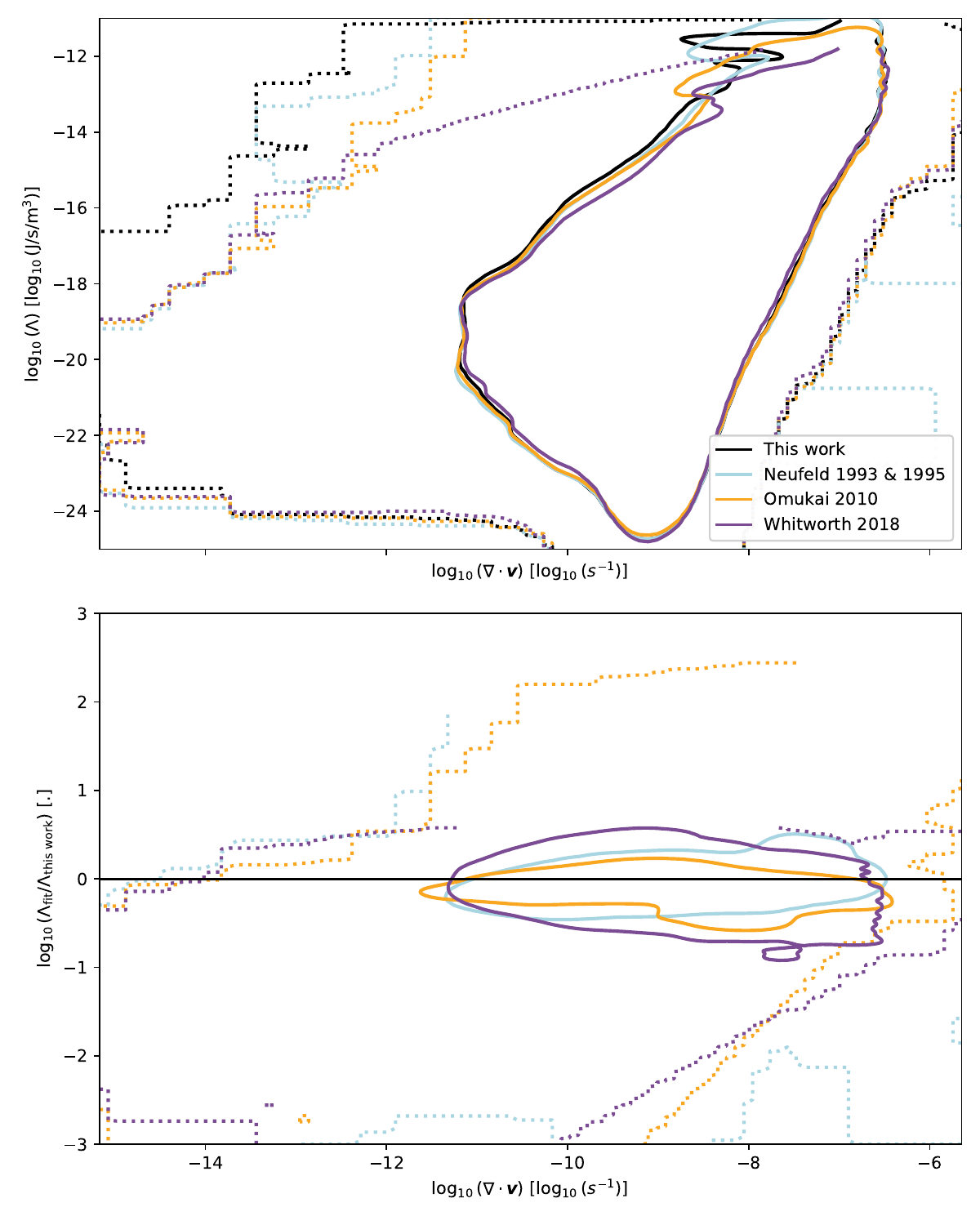}
	\caption{Contour plots of the cooling rate versus the velocity divergence, smoothed using a Gaussian kernel. The solid line contains $99\%$ of the data points. The dotted line denotes the outer contour where data is present. Bottom: contour plot of the relative differences between the fits and our computed cooling rate.}
	\label{figure: contour plot velocity divergence}
\end{figure}

\begin{figure}
	\includegraphics[width=\linewidth]{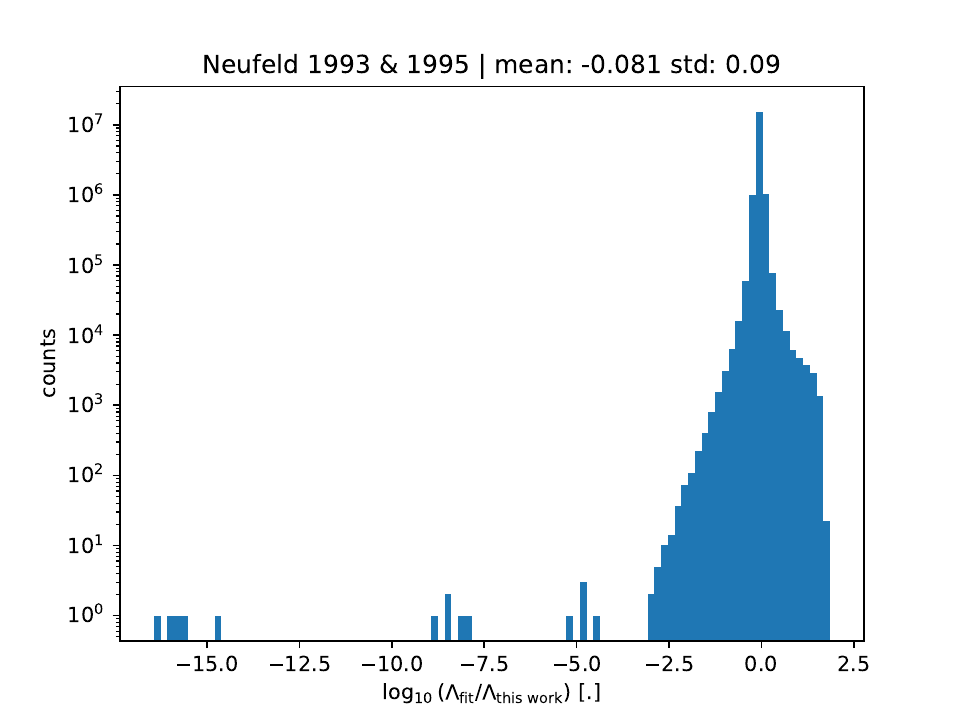}
	\caption{Histogram of relative differences between the reference results and the cooling fit result calculated using the combined Neufeld prescriptions \citep{neufeld_radiative_1993, neufeld_thermal_1995}, in log space. Note that the y-axis is also in log space to show the outliers.}
	\label{figure: residuals fit neufeld}
\end{figure}

\begin{figure}
	\includegraphics[width=\linewidth]{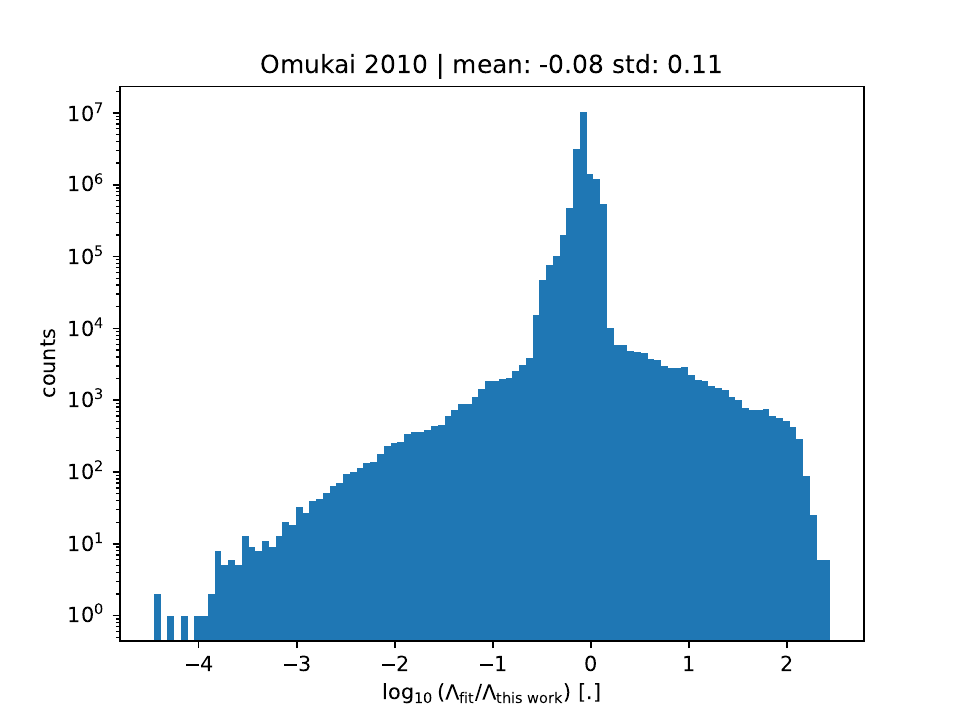}
	\caption{Histogram of relative differences between the reference results and the cooling fit result calculated using the \protect\cite{omukai_low-metallicity_2010} prescription, in log space. Note that the y-axis is also in log space to show the outliers.}
	\label{figure: residuals fit omukai}
\end{figure}

\begin{figure}
	\includegraphics[width=\linewidth]{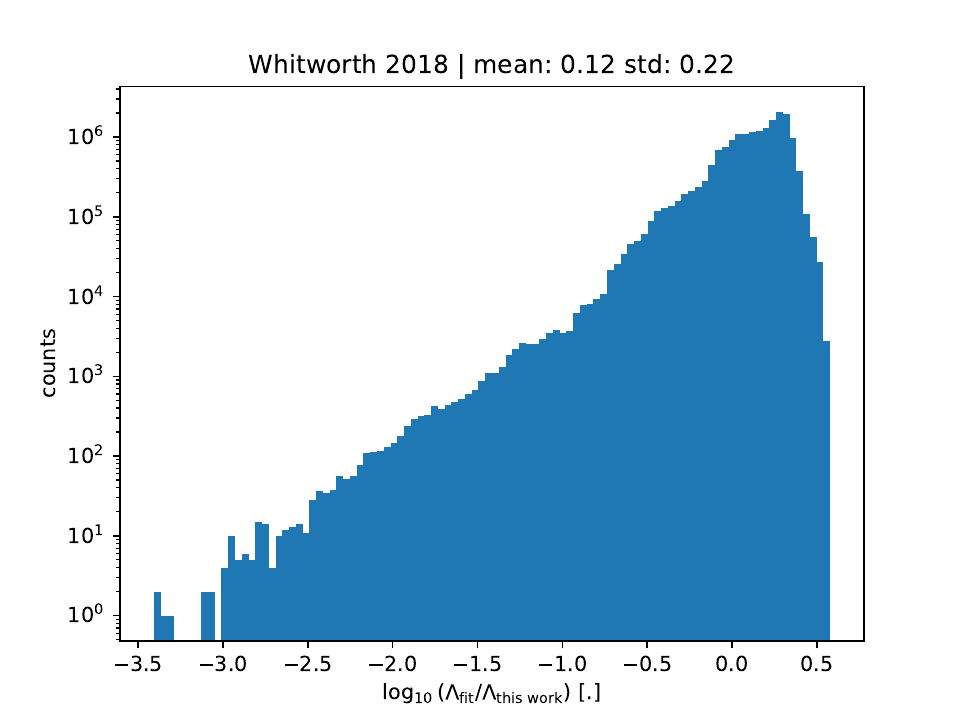}
	\caption{Histogram of relative differences between the reference results and the cooling fit result calculated using the \protect\cite{whitworth_simple_2018} prescription, in log space. Note that the y-axis is also in log space to show the outliers.}
	\label{figure: residuals fit whitworth}
\end{figure}

\begin{figure}
	\includegraphics[width=\linewidth]{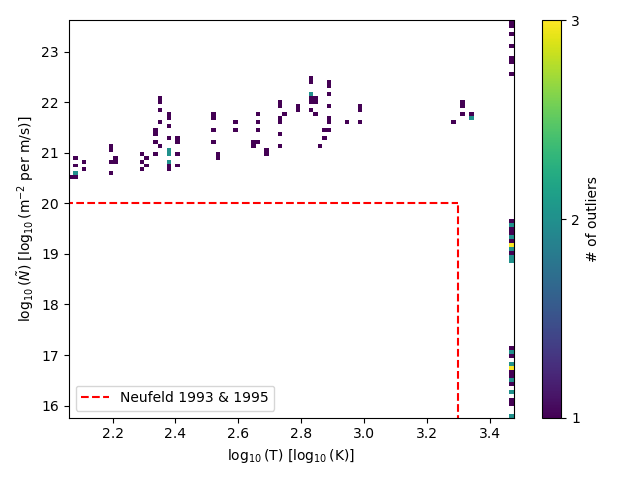}
	\caption{Location of the outliers for the computed cooling rates, when applying the combined Neufeld prescription \citep{neufeld_radiative_1993, neufeld_thermal_1995}. All outliers lie outside the original intended parameter range for this prescription.}
	\label{figure: outliers fit neufeld}
\end{figure}

\begin{figure}
	\includegraphics[width=\linewidth]{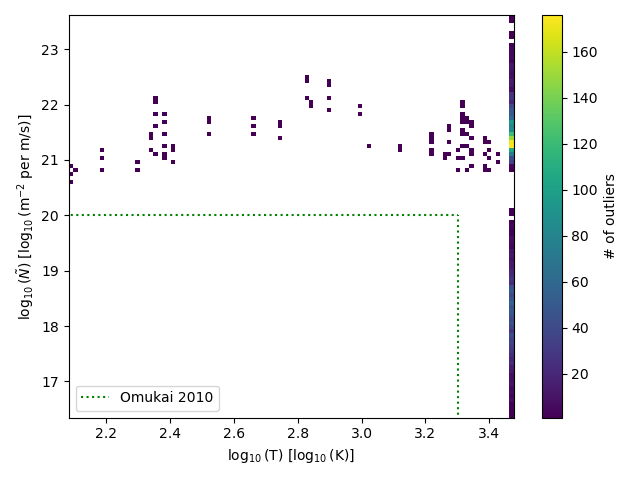}
	\caption{Location of the outliers for the computed cooling rates, when applying the \protect\cite{omukai_low-metallicity_2010} prescription. All outliers lie outside the original intended parameter range for this prescription.}
	\label{figure: outliers fit omukai}
\end{figure}

\begin{figure}
	\includegraphics[width=\linewidth]{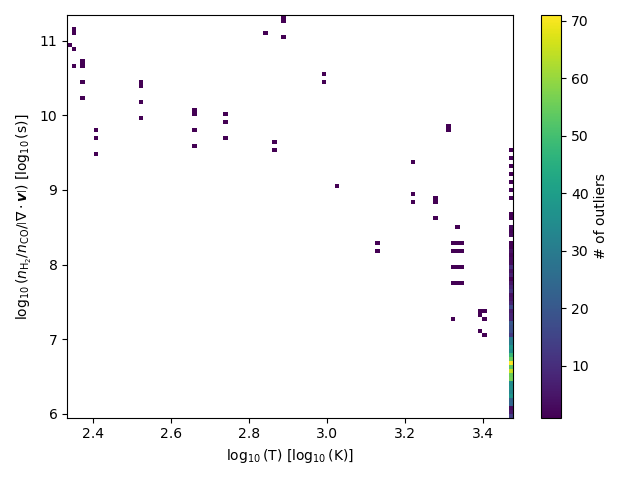}
	\caption{Location of the outliers for the computed cooling rates, when applying the \protect\cite{whitworth_simple_2018} prescription. All outliers lie outside the original intended parameter range for this prescription.}
	\label{figure: outliers fit whitworth}
\end{figure}

None of the existing prescriptions offer a parameter range which encompasses the entire parameter range of our models (see Figs. \ref{figure: parameter_bounds_neufeld_omukai}, \ref{figure: parameter_bounds_whitworth}).
Therefore, we do not expect the cooling rate prescriptions to be appropriate for the entire parameter range of our models. However, all of them offer a reasonable proxy for the computed cooling rate (see Figs. \ref{figure: contour plot co density}, \ref{figure: contour plot temperature}, \ref{figure: contour plot velocity divergence}), especially when evaluated inside their intended parameter range (see Appendix \ref{appendix: existing cooling prescriptions within and outside their parameter range}). We show the relative differences with the computed cooling rate in Figs \ref{figure: residuals fit neufeld}, \ref{figure: residuals fit omukai} and \ref{figure: residuals fit whitworth} respectively. From these figures, we observe that on average, the combined fit from \cite{neufeld_radiative_1993, neufeld_thermal_1995} performs best, with a mean relative difference of $-0.08$ dex, however, some very far away outliers exist. Similarly, the \cite{omukai_low-metallicity_2010} prescription slightly underestimates the cooling rate, with a factor of $0.08$ dex and has a standard deviation on the error of about $0.11$ dex. Finally, the fit from \cite{whitworth_simple_2018} slightly overestimates the cooling rate on average (by $0.12$ dex), and has a higher standard deviation of about $0.22$ dex. 
To complete our analysis, we check the number of outliers for each fit prescription, which we define as
\begin{align}\label{eq: outlier definition}
	\left|\log_{10}(\Lambda_{\text{fit}}/\Lambda_{\text{this work}})\right|>2.
\end{align}
Each outlier therefore corresponds to a prediction for the cooling rate which is off by at least two orders of magnitude. The prescriptions of \cite{neufeld_radiative_1993, neufeld_thermal_1995} combined, \cite{omukai_low-metallicity_2010}, and \cite{whitworth_simple_2018} respectively have $145$, $2622$, and $1015$ outliers when applied to our complete data set. In Figs. \ref{figure: outliers fit neufeld}, \ref{figure: outliers fit omukai}, and \ref{figure: outliers fit whitworth}, we have plotted the locations of the outliers in our parameters space. Unsurprisingly, the outliers all lie outside the intended parameter regimes of the fit prescriptions, in the more challenging high temperature and/or high density regime.
However, as any outlier present in the fit function is one too many, we will create a new fit in Section \ref{section: fit all}.


\subsection{New cooling rate prescription}\label{section: fit all}

We fit the data according to a polynomial fit function in logarithmic space. As we do not know how many fit parameters we require a priori, we allow the number of polynomial terms per input parameter to be variable. In the fit, we use $N_1$ terms for the temperature $T$, $N_2$ terms for the H$_2$ density, $N_3$ terms for the abundance ratio $n_{\text{CO}}/n_{\text{H}_2}$, $N_4$ terms for $\tilde{N}(\text{CO})$, resulting in the following formula for the $\log_{10}$ of the cooling rate $\Lambda$:
\begin{align}
	\log_{10}(\Lambda)_{\text{fit}} &= a + \sum_{i=1}^{N_1} b_i(\log_{10}(T))^{i} + \sum_{i=1}^{N_2} c_i(\log_{10}(n_{\text{H}_2}))^{i}\\
	&+\sum_{i=1}^{N_3} d_i\left(\log_{10}\left(\frac{n_{\text{CO}}}{n_{\text{H}_2}}\right)\right)^{i} + \sum_{i=1}^{N_4} e_i\left(\log_{10}\left(\frac{n_{\text{CO}}}{\left|\nabla\cdot \boldsymbol{v}\right|}\right)\right)^{i}.\nonumber
\end{align}
In this, $a$ is a constant term, and $b_i, c_i, d_i, e_i$ are constants (see Eq. \ref{eq: fit parameters full fit}) with the number of parameters per quantity $N_i$ to be determined. The input terms for the fits are all in SI units. Note that we chose the abundance ratio $n_{\text{CO}}/n_{\text{H}_2}$ instead of the CO number density in our fit. This is because we have modeled the CO number density to be proportional to the H$_2$ number density, and is therefore equal to the log H$_2$ number density plus a constant in log space. Therefore, also using a CO density term would result in an ill-defined fit function. The lower bound of $N=[N_1, N_2, N_3, N_4]$ we consider, is $N=[1,1,1,0]$. The upper bound we consider is $N=[4,4,4,4]$. We limit the order of the polynomial terms to $4$, in order to limit any issues with numerical precision, as the optimal fit coefficients would significantly increase when more higher order terms are included.
As the fit coefficients try to cancel each other out, small perturbations on the coefficients (by e.g. rounding), can result in large perturbations on the fit result. We therefore do not round our obtained fit coefficient values in this paper.
Furthermore, in order to further reduce the values of the obtained fit coefficients, we add a regularization term for the coefficients to the least-squares fitting we use, defining the loss function $\mathcal{L}$ to be:
\begin{align}\label{eq: loss function regularization}
\mathcal{L} &=  \sum_{\text{data points}}\left(\log_{10}(\Lambda)_{\text{fit}} - \log_{10}(\Lambda)_{\text{data}}\right)^2 \\
&+ r^2\left(a^2 + \sum_{i=1}^{N_1} \left((i+1)b_i\right)^2 + \sum_{i=1}^{N_2} \left((i+1)c_i\right)^2\right. \nonumber \\
&+ \left. \sum_{i=1}^{N_3} \left((i+1)d_i\right)^2 + \sum_{i=1}^{N_4} \left((i+1)e_i\right)^2\right) \nonumber
\end{align}

in which the coefficient of the regularization term $r$, is taken to be $1$.
We start from the most complicated configuration $N = [4,4,4,4]$, and removed polynomial terms, starting from the last parameter, unless the mean fit error would change significantly, by more than $1\%$. 
Afterwards, we repeat this procedure for all other parameters, from (second to) last to first. 
The final configuration uses $N=[3,4,1,2]$, with parameter values

\begin{align}
a &= -1.74249389\cdot 10^1\label{eq: fit parameters full fit}\\
\boldsymbol{b} &= [1.08678135,  -6.83716540\cdot 10^{-2}, 3.08211028\cdot 10^{-2}]\nonumber\\
\boldsymbol{c} &= [-6.13253761,  1.01649680, -5.32517202\cdot 10^{-2},\nonumber\\&
9.04689294\cdot 10^{-4}]\nonumber\\
\boldsymbol{d} &= [9.61193214\cdot 10^{-1}]\nonumber \\
\boldsymbol{e} &= [2.23296121\cdot 10^{-1}, -6.30372565\cdot 10^{-3}]\nonumber.
\end{align}
The fit residual has a standard deviation of $0.034$ dex.

\begin{figure}
	\includegraphics[width=\linewidth]{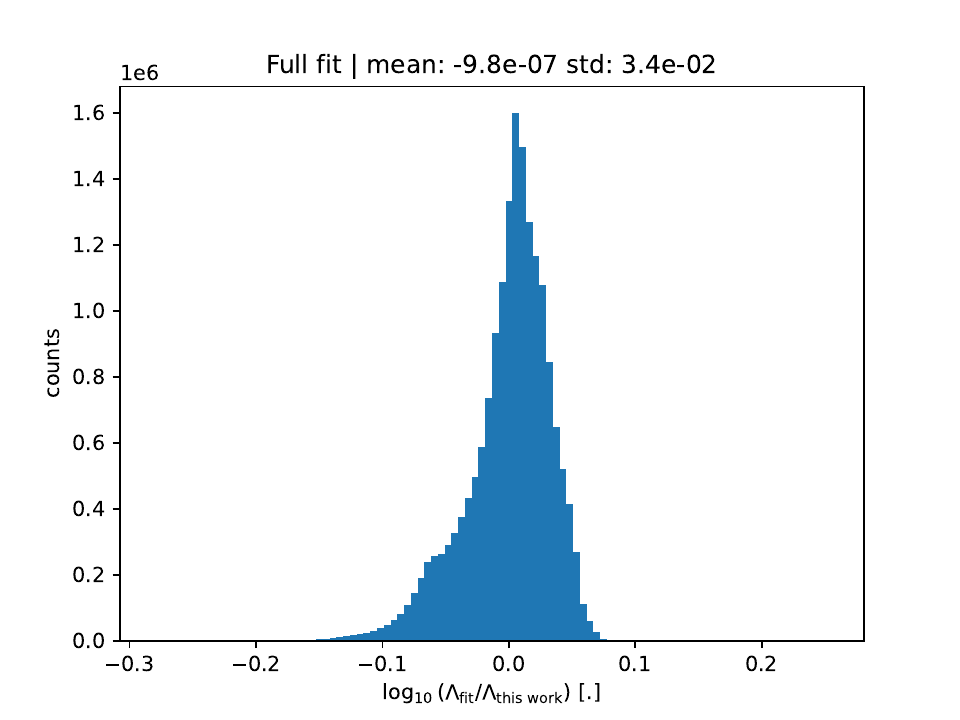}
	\caption{Histogram of relative differences between the reference results and the cooling fit of this paper for the entire parameter range.}\label{figure: residuals fit full}
\end{figure}

\begin{figure}
	\includegraphics[width=\linewidth]{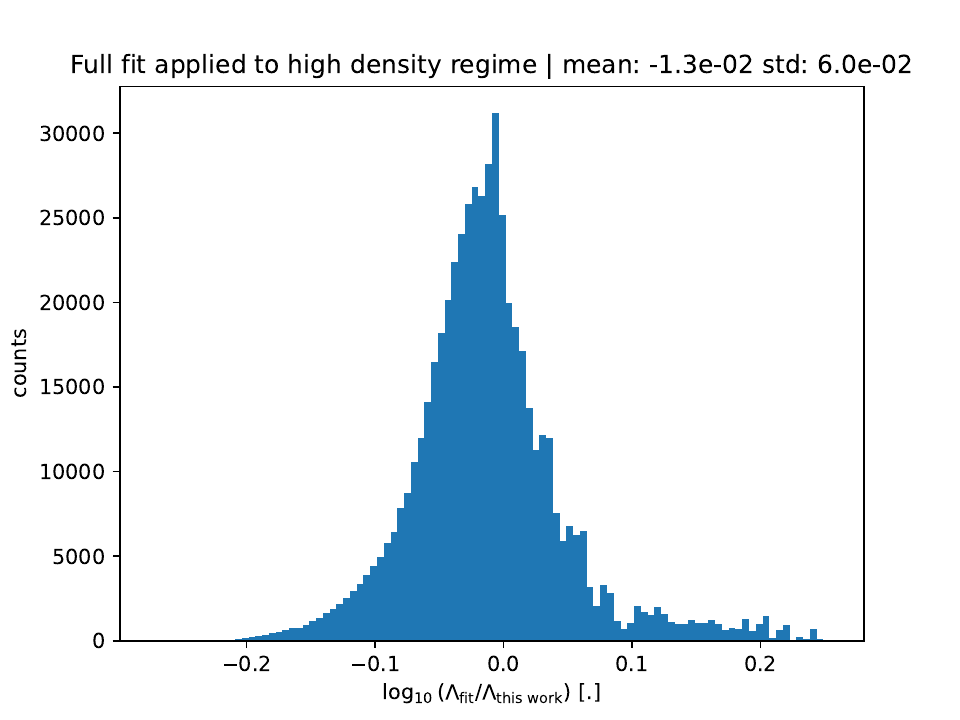}
	\caption{Histogram of relative differences between the reference results and the cooling fit of this paper for the entire parameter range, evaluated in the high density regime $\log_{10}(n_\text{CO})>10$.}\label{figure: residuals fit full high density}
\end{figure}

\begin{figure*}
	\includegraphics[width=\linewidth]{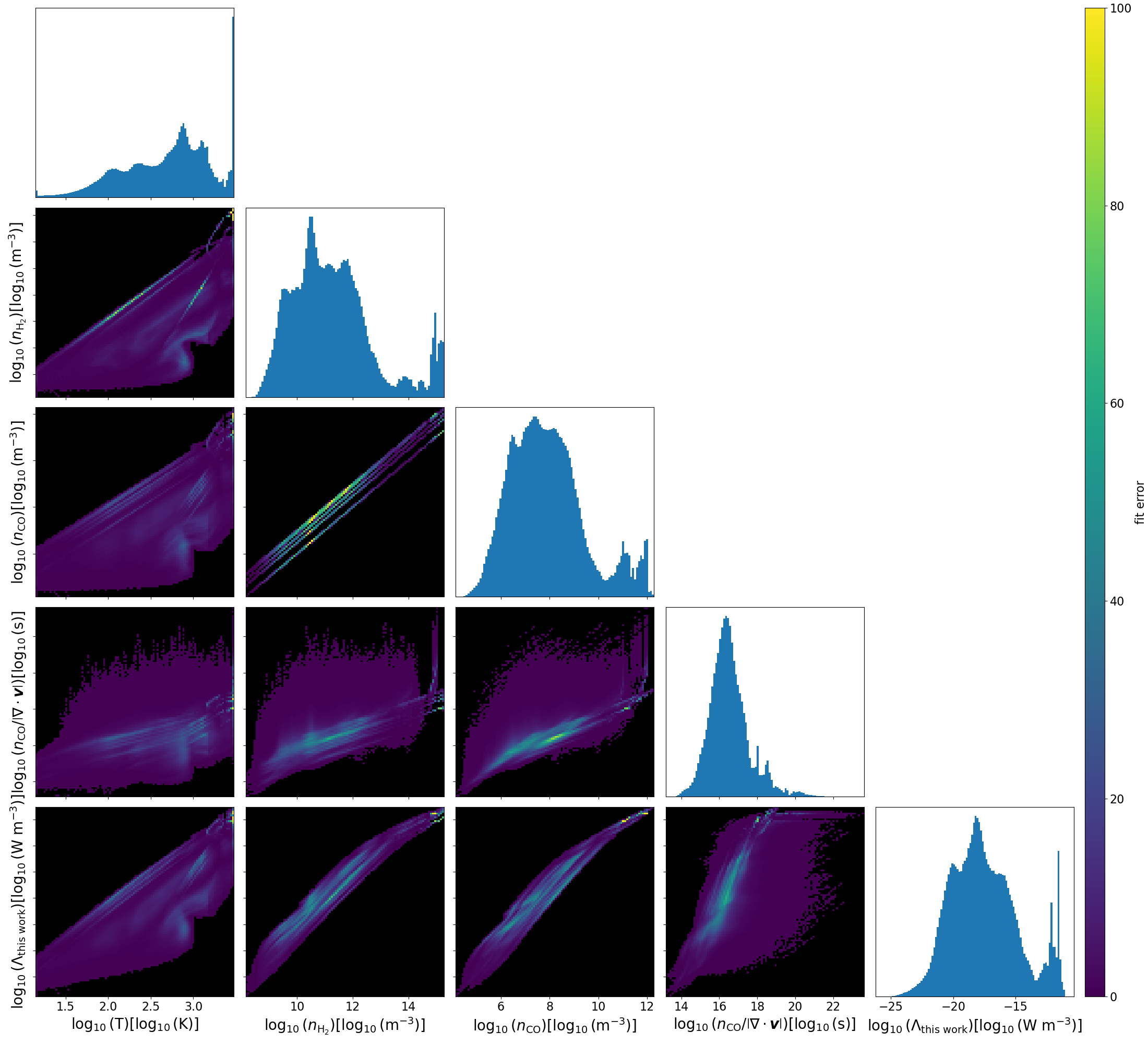}
	\caption{Corner plot of the fit error corresponding to the full parameter space fit of Section \ref{section: fit all}, in which we define the fit error as $(\log_{10}(\Lambda_{\text{fit}})-\log_{10}(\Lambda_{\text{this work}}))^2$. The parameter space without data is colored black. 
	}\label{figure: cornerplot loss function}
\end{figure*}

Upon further inspection of the residual, we found that the general fit performs worse in the highest density regions (see Fig. \ref{figure: residuals fit full high density}). The parameter region for which $\log_{10}(n_\text{CO})>10 \log_{10}(\text{m}^{-3})$ contains about $3\%$ of all model points, but is fitted slightly worse than the lower density regions. As there are $2$ orders of magnitude less data points in the former regime compared to the latter, the former will be fitted worse. To illustrate this, we create a corner plot using the fit residuals, squared\footnote{Such that we mimic the L2 norm loss function used for least squares fitting.}, shown in Fig. \ref{figure: cornerplot loss function}. This figure shows in which part of the parameter space the fit performs worst, and is not corrected for data point density, similar to the actual fitting procedure. When applying our fit to only the high density regime, we obtain a standard deviation of the residual of $0.06$ dex, which is worse than the general fit, but in our opinion does not necessitate a separate fit for the high density regime.


\subsection{Estimating the fit error}\label{section: estimating fit error}

When using least-squares minimization, one obtains by default the optimal fit parameters $\boldsymbol{w}$, but one can also obtain a covariance matrix $\Sigma_{\boldsymbol{w}}$ for the obtained fit parameters, which can be used for estimating the uncertainty on the fit parameters. In this paper, we use the function \texttt{scipy.optimize.curve\_fit} from \textsc{Scipy} \citep{virtanen_scipy_2020} to compute both simultaneously. To derive uncertainties for the fit, we use

\begin{align}
\Sigma_f = X \Sigma_{\boldsymbol{w}} X^T,
\end{align}
in which $\Sigma_f$ is the covariance matrices for the fit function value. In this, $X$ denotes the $(N_{\text{evaluations}}\times N_{\text{fit parameters}})$ design matrix, in which each row corresponds to the evaluation of a single data point at all individual polynomial terms, such that the fit result $f(x_i)$ of a single data point $x_i$ can be evaluated using matrix multiplication, i.e.
\begin{align}
f(x_i) = X_{i} \boldsymbol{w},
\end{align}
in which $X_{i}$ denotes the $i$-th row of the matrix $X$. We can now estimate the residuals by calibrating the fit uncertainty to the variance of the fit residual, using
\begin{align}
\Sigma_r = c^2 \Sigma_f
\end{align}
in which $c$ is given by
\begin{align}
c = \frac{\sigma}{\sqrt{E(\Sigma_{f, ii})}}
\end{align}
and $\sigma$ is the standard deviation on the fit residual and $\sqrt{E(\Sigma_{f, ii})}$ is the square root of the average of the diagonal elements of $\Sigma_f$, evaluated on all data points used for the fit\footnote{If one is not particularly interested in the covariance, but just aims to use $\Sigma_r$ for estimating the pointwise error, one does not need to compute the off-diagonal elements of $\Sigma_f$.}. To shorten notation, we now define $\sigma_{\text{fit}}$ to be the square root of the diagonal of $\Sigma_r$. Both $\Sigma_{\boldsymbol{w}}$ and $c$ can be found in the online supplementary data.

\begin{figure}
	\includegraphics[width=\linewidth]{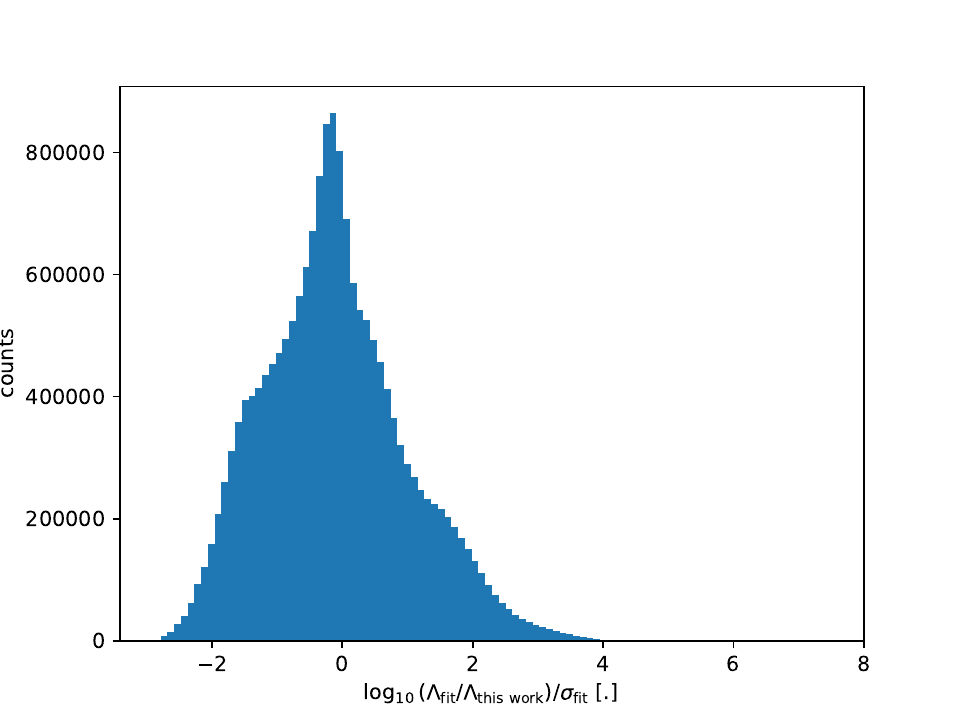}
	\caption{Histogram of the fit residual, divided by the calibrated fit error, calculated using the fit based on the entire parameter range.}\label{fig: fit error histogram}
\end{figure}

\begin{figure}
	\includegraphics[width=\linewidth]{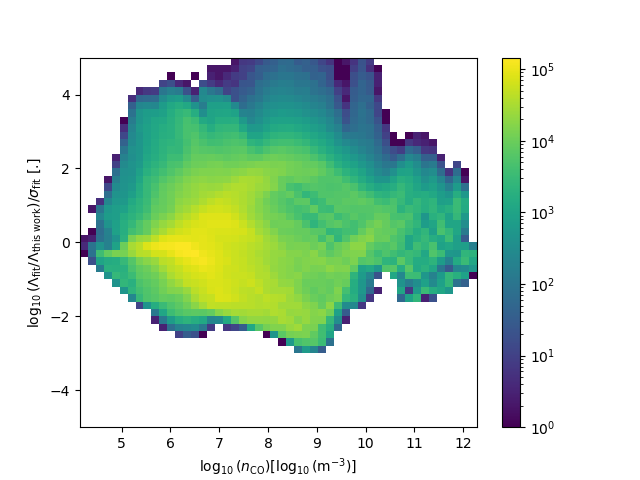}
	\caption{2D histogram of the fit residual, divided by the calibrated fit error, plotted against the CO density, using a logarithmic scale for the color bar.}\label{fig: fit error hist2D}
\end{figure}

To check whether this calibrated fit uncertainty agrees with the actual residual, we calculate this estimate for our fit in this paper (Eq. \ref{eq: fit parameters full fit}), for all data points. We find that $67\%$ of the fit residuals lie between $\pm 1\sigma_{\text{fit}}$ and $96\%$ lie between $\pm 2\sigma_{\text{fit}}$ (see Fig. \ref{fig: fit error histogram}). From Fig. \ref{fig: fit error hist2D}, we notice that in the intermediate density regime, $7<\log_{10}\left(n_{\text{CO}}\right)<10$, the calibrated fit error $\sigma_{\text{fit}}$ seems to underestimate the residual, as the residual shows a two-peaked distribution in this part of the parameter space. Conversely, in the very low density regime, $\log_{10}\left(n_{\text{CO}}\right)<5$, $\sigma_{\text{fit}}$ overestimates the residual. We conclude that the calibrated fit error $\sigma_{\text{fit}}$ can be used as proxy for (the width of the distribution of) the fit residual, however, it might not be fully accurate when looking at specific regions of the parameter space.

\section{Model perturbations}\label{section: model perturbations}

In Section \ref{section: estimating fit error}, we have computed uncertainties on the accuracy of the obtained fit. However, the physics included in the hydrodynamics simulations which we used for calculating the CO cooling rates, are not entirely complete, missing for example the effect of CO cooling itself. To estimate how much the cooling rates may change with a slightly different physics description, we study the influence of using slightly perturbed parameter values on the cooling rate. For this we use \textsc{Magrittetorch}\footnote{Available at \url{https://github.com/Magritte-code/Magritte-torch}}, a port of \textsc{Magritte} \citep[see e.g.][]{de_ceuster_simulating_2022} to \textsc{Pytorch} \cite{paszke_pytorch_2019} \footnote{\textsc{Magrittetorch} is currently based on version 0.7.2 of \textsc{Magritte}.}. We have built this code on top of a machine learning library to ensure that the resulting models are fully differentiable. This allows us to track a gradient throughout the non-linear radiative transfer calculations. Starting from a regular radiative transfer model, we can use forward gradients to add a perturbation term on the model input, which is propagated throughout the entire NLTE computation, eventually giving us a gradient on the computed cooling rates.

To get a global idea of how the cooling rates change, depending on perturbations of the model input, we rerun the last snapshot of each AGB binary model used in this paper, for a CO/H$_2$ abundance ratio of $5\cdot 10^{-4}$. We successively apply perturbations $\delta$ of $10\%$ on the temperature, CO density and velocity. Note that the exact percentage of the perturbations does not matter, as we obtain a linear approximation for perturbation through evaluating the gradient, and we will rescale the obtained perturbation on the cooling rate by the given percentage on the input gradient. In this way, we obtain the change in (log) cooling rate per (log) input parameter. For example, when applying this for the input temperature, we obtain
\begin{align}
\alpha = \frac{\Lambda(T+\delta T)}{\Lambda(T)/\delta} = \frac{\Lambda(T+\delta T)}{\Lambda(T)}\frac{T}{\delta T} \simeq \frac{\partial \log(\Lambda)}{\partial \log(T)},
\end{align}
in which $\alpha$ is the computed value. By integrating and exponentiating both sides, we obtain a (local) scaling relation for the cooling rate
\begin{align}
\Lambda \sim T^\alpha.
\end{align}
A similar argument can be made to obtain scaling relations for the other input parameters.

\begin{figure}
	\includegraphics[width=\linewidth]{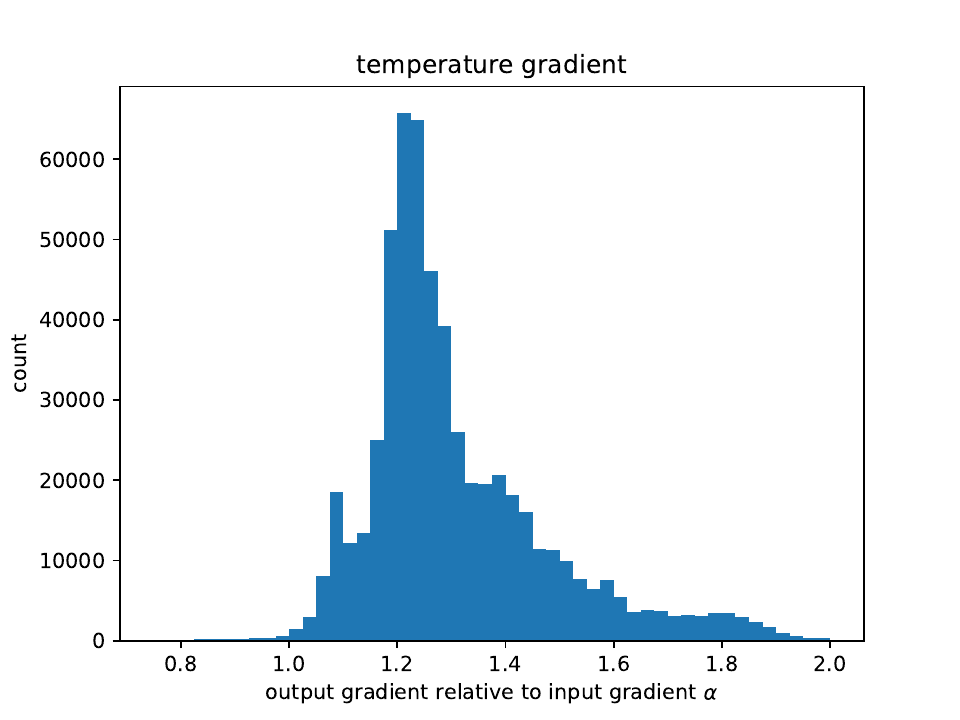}
	\caption{Histogram of the impact of a temperature perturbation on the cooling rate in the low density regime.}\label{fig: cooling gradient temp low density}
\end{figure}

\begin{figure}
	\includegraphics[width=\linewidth]{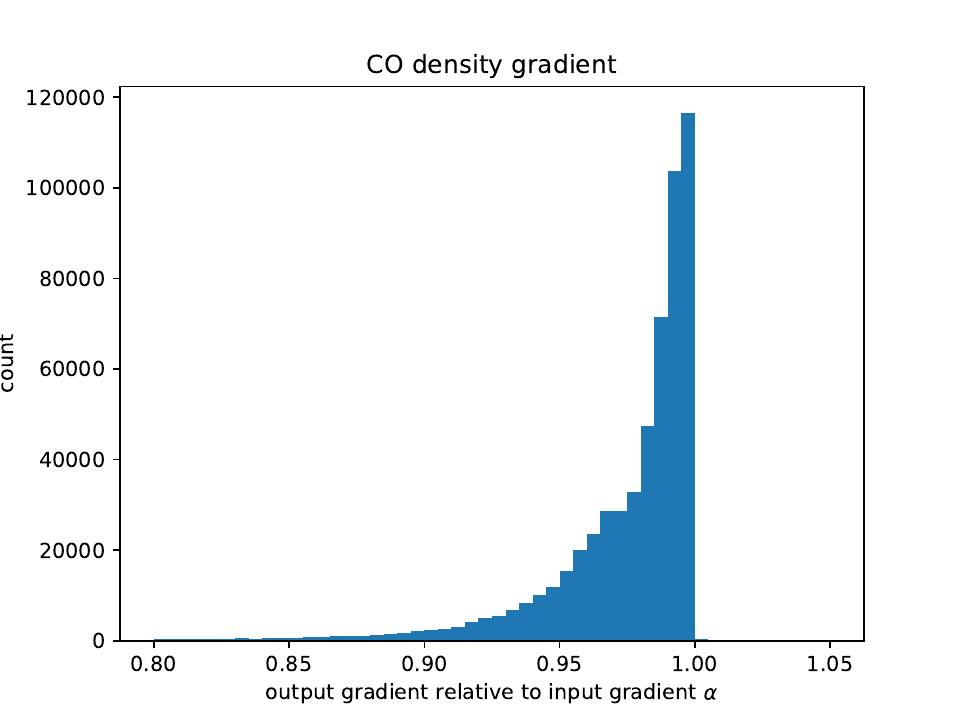}
	\caption{Histogram of the impact of a CO density perturbation on the cooling rate in the low density regime.}\label{fig: cooling gradient CO low density}
\end{figure}

\begin{figure}
	\includegraphics[width=\linewidth]{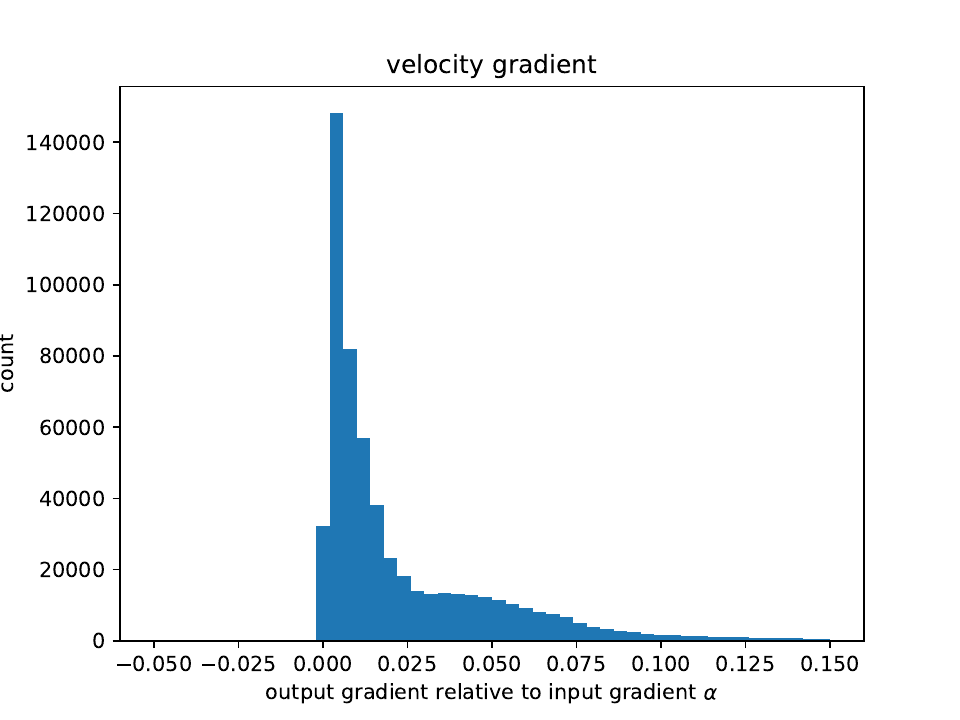}
	\caption{Histogram of the impact of a perturbation of the velocity field on the cooling rate in the low density regime.}\label{fig: cooling gradient vel low density}
\end{figure}

We start by first exploring the impact of the input parameters on the low density regime, $\log_{10}(n_{\text{CO}})<10$. In Fig. \ref{fig: cooling gradient temp low density}, we see that an increase in temperature in general leads to a (super)linear increase in cooling. During the analysis, however, we have noticed some outliers. Data points with temperatures above $3000$K are treated inconsistently during the radiative transfer calculation, due to collisional data being limited to $\leq 3000$K. In our code, temperatures above $3000$K still impact the line width, but no longer influence the collision rates $C_{ij}$. Therefore, we have removed these data points from our analysis for this section. Next, we plot the impact of a perturbation for the CO density on the cooling rate in Fig. \ref{fig: cooling gradient CO low density}. We find that the cooling rate depends linearly on the density, which is as we expect for the optically thin regime. Finally, in Fig. \ref{fig: cooling gradient vel low density}, we find that the cooling rate does almost not depend on the velocity field in this regime, which can be expected based on the argumentation in Section \ref{section: existing cooling prescriptions}.

\begin{figure}
	\includegraphics[width=\linewidth]{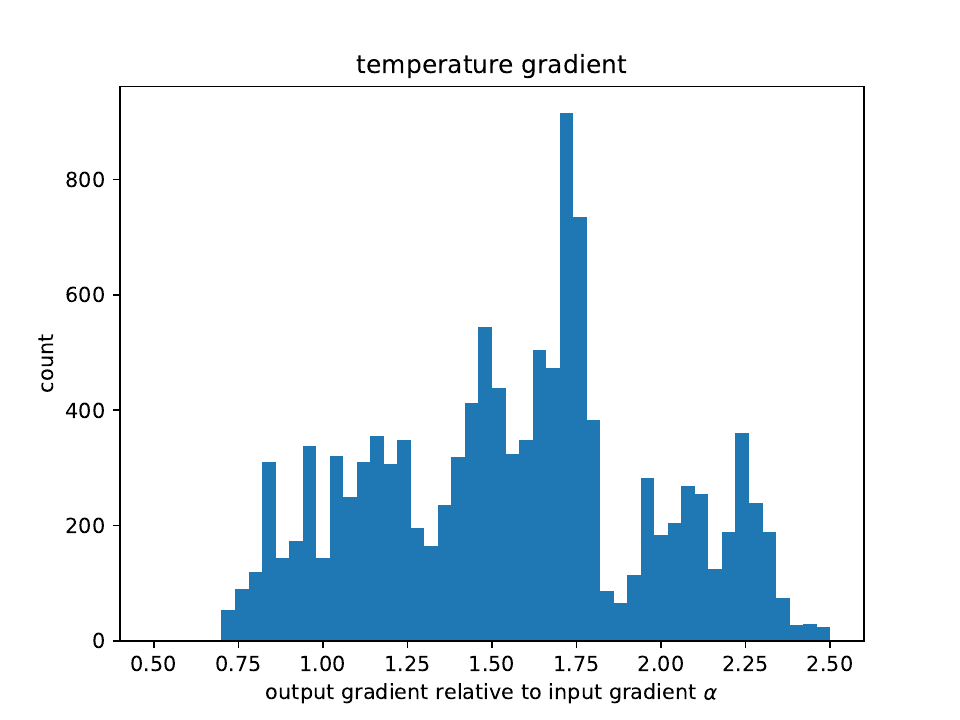}
	\caption{Histogram of the impact of a temperature perturbation on the cooling rate in the high density regime.}\label{fig: cooling gradient temp high density}
\end{figure}

\begin{figure}
	\includegraphics[width=\linewidth]{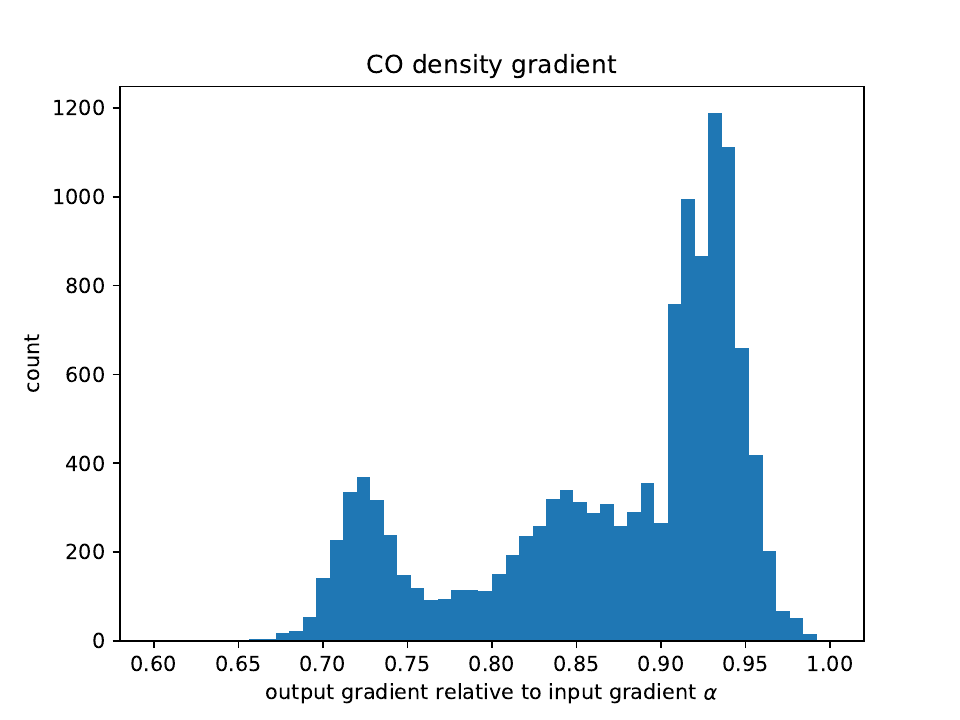}
\caption{Histogram of the impact of a CO density perturbation on the cooling rate in the high density regime.}\label{fig: cooling gradient CO high density}
\end{figure}

\begin{figure}
	\includegraphics[width=\linewidth]{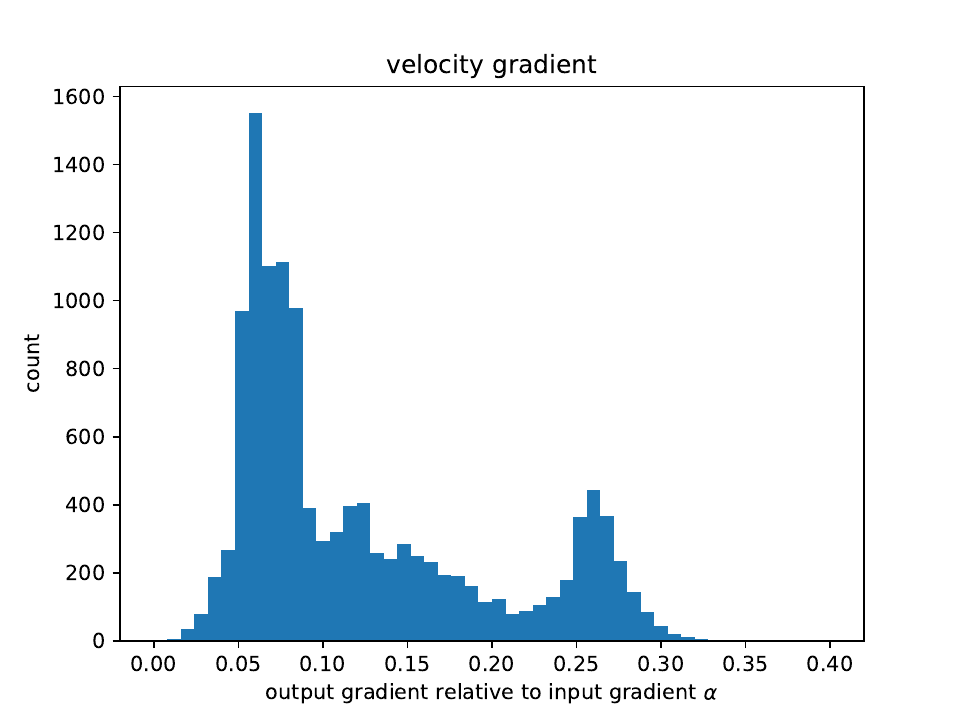}
	\caption{Histogram of the impact of a perturbation of the velocity field on the cooling rate in the high density regime.}\label{fig: cooling gradient vel high density}
\end{figure}

\begin{figure}
	\includegraphics[width=\linewidth]{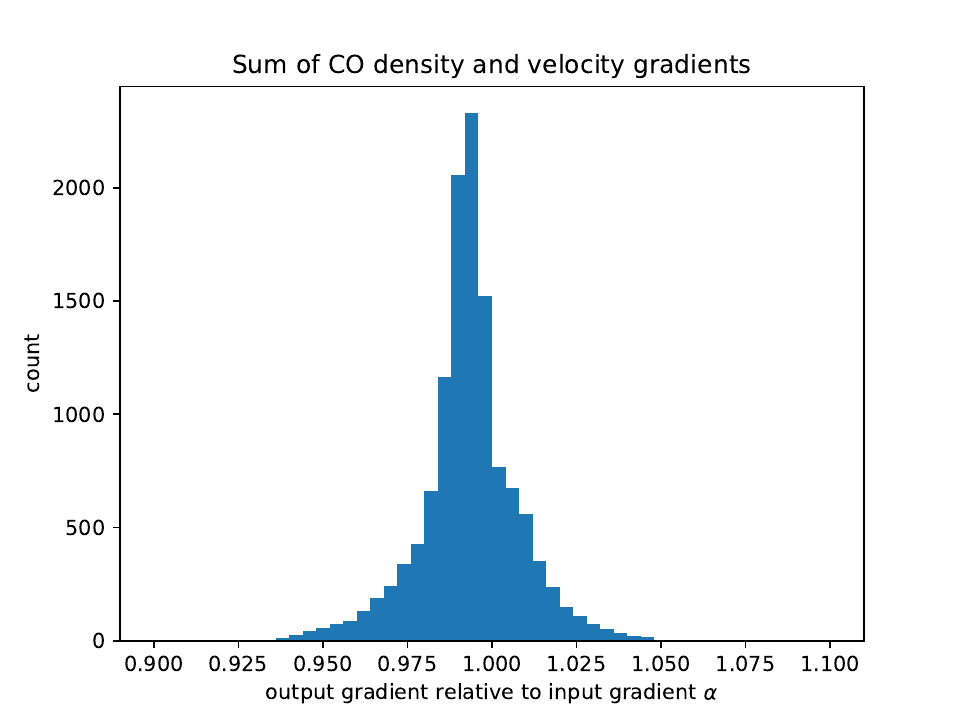}
	\caption{Histogram of the sum of impacts $\alpha$ of a perturbation on the CO density and the velocity field on the cooling rate in the high density regime.}\label{fig: cooling gradient CO and vel high density}
\end{figure}

For the high density regime, $\log_{10}(n_{\text{CO}})>10$, we plot the cooling rate dependence on the same input parameters in Figs. \ref{fig: cooling gradient temp high density}, \ref{fig: cooling gradient CO high density} and \ref{fig: cooling gradient vel high density}. 
Here, we find similar results for the cooling rate scaling as in the low density regime, but with a slightly weaker dependence on the CO density, and higher dependence on the velocity gradient. The obtained temperature gradient is slightly higher on average. Curiously, when adding the CO density gradient and velocity gradient together, see Fig. \ref{fig: cooling gradient CO and vel high density}, we obtain values for the sum of $\alpha$ close to $1$, which implies $\Lambda\sim n_{\text{CO}}^{\alpha_\text{CO}}\left|\nabla\cdot \boldsymbol{v}\right|^{1-\alpha_\text{CO}}$. This indicates that when the cooling becomes less impacted by increasing density (in the optically thick regime), the velocity gradient becomes increasingly more important. This makes sense, as the velocity gradient allows for light to escape in wider frequency bands, up to a certain maximum of total light which can be emitted in the optically thin regime (when $\Lambda\sim n_\text{CO}$).
 
We note that most scaling relations found in this section can be derived from analytical approximations \citep[see e.g.][]{whitworth_simple_2018}, but we hereby confirm these to be valid for radiative transfer simulations in the complex AGB environment. However, the exact power laws of the scaling laws might not entirely match. In particular, \cite{whitworth_simple_2018} predicts for the temperature scaling in the high density regime that $\Lambda \sim T^{4}$, and for the velocity gradient scaling $\Lambda \sim \left|\nabla\cdot \boldsymbol{v}\right|$. Our results do not come close to the aforementioned temperature scaling and velocity gradient scaling. We must note however, that these discrepancies might arise from be due to our high density model points not being dense enough, as the locally emitted radiation might not yet be completely limited by the velocity field.

\section{Limitations}\label{section: limitations}
The cooling fits we propose in this paper come with some limitations to the regime they are applicable to. Aside from the parameter limits described in Table \ref{table: SPH parameter limits}, we list some shortcomings of our radiative transfer modeling and estimate their impact on the obtained rotational CO line cooling rates.

Our radiative transfer models in this paper do not include the intensity contribution from the star. We have done this to allow our results to be comparable with literature prescriptions for ISM-like conditions in Section \ref{section: existing cooling prescriptions}. As the regions just outside the star are approximately in LTE conditions, and have temperatures similar to the effective temperature of the star, we do not think this omission has a significant impact on the radiation field in the models. However, an extra intensity contribution from the star might cause the closest regions to heat instead of cool. We hereby note that we not encountered any net radiative heating in any of our \textsc{Magritte} models.

We have simulated NLTE line radiative transfer using the first 75 rotational levels of CO (v=0), using the collisional rates \citep[from][]{yang_rotational_2010}, extrapolated to higher rotational transitions using the procedure described in \cite{schoier_atomic_2005} (see details in Appendix \ref{appendix: CO collisional rate extrapolation}). We have therefore implicitly imposed an upper bound on the model temperature of $3000$K, when simulating NLTE radiative transfer. Correspondingly, we also imposed an upper bound of $3000$K on the model temperatures used for the fitting functions. Finally, we have neglected any vibrational transitions, given that we do not have the required collisional data for simulating these vibrational transitions in NLTE (see Section \ref{section: general model setup}). We found that by increasing the number of modeled rotational transitions, the net cooling rate generally increases at higher temperatures (see Appendix \ref{appendix: CO collisional rate extrapolation}). Based on this, we expect that including vibrational levels would slightly increase the computed cooling rate $\Lambda$ for temperatures starting at around $750$ K, which is an estimate for the temperature at which the first vibrational state would be significantly populated.


\section{Conclusion}
In this paper, we provide CO cooling rates for the AGB environment. For this, we create various models of the AGB binary environment in Section \ref{section: CO cooling rates for AGB outflows}, to which we apply NLTE line radiative transfer. We first compare the computed cooling rates with some cooling rate prescriptions for the ISM, which we describe in Section \ref{section: existing cooling prescriptions}. It turns out that all of these prescriptions are based on different parameter ranges than we have in our AGB models. We found that the combined cooling rate prescription from \cite{neufeld_radiative_1993, neufeld_thermal_1995} works best. This prescription slightly underestimates the cooling rate by $0.08$ dex on average, and has a standard deviation on the error of $0.09$ dex. However, outside their intended parameter regime, all literature prescriptions considered in this paper can show significant errors on order of multiple dex.
To improve upon this for the AGB parameter regime, we propose a polynomial fit for the cooling rate in Section \ref{section: fit all}. We find that using a single fit for the entire parameter range, results in a bit worse fit in the high density regime, with a standard deviation on the error of $0.06$ dex, compared to $0.03$ dex in general, but still performs adequate enough to not need a seperate fit for the high density regime.
The cooling rates obtained in this paper valid are for the AGB outflow, with temperatures limited to $<3000$ K, H$_2$ densities between $1.5\cdot 10^8 \text{ m}^{-3}$ and $1.9\cdot 10^{15} \text{ m}^{-3}$, CO$/$H$_2$ ratios between $10^{-4}, 10^{-3}$ and $\tilde{N}$ between $1.5\cdot 10^{13}$ and $4.2\cdot 10^{23}$ s m$^{-3}$.
Finally, we numerically verified in Section \ref{section: model perturbations} some analytic scaling relations for the cooling rate, both in the low and high density regimes in the AGB environment. In particular, we confirm that the velocity gradient $\left|\nabla\cdot\boldsymbol{v}\right|$ has little impact on the cooling rate in the low density regime, but matters slightly more in the higher density regime, where the impact of the CO density drops.

\section*{Acknowledgments}
T. Ceulemans has been supported by the Research Foundation - Flanders (FWO), grant 1166724N.
F. De Ceuster has been supported by the FWO, grants I002123N and 1253223N.
O. Vermeulen has been supported by the FWO, grant 1173025N.
L. Decin has been supported by the FWO, grants G099720N, G0B3823N and the KU Leuven methusalem grant METH/24/012. 
The authors express their thanks to J. Malfait for providing a \textsc{Phantom} model which has been used in this paper.

\section*{Conflict of Interest}
The authors declare no conflicts of interest.


\section*{Data Availability}

\textsc{Magritte} is an open-source software library available at \url{https://github.com/Magritte-code/Magritte}. \textsc{Magrittetorch} is an open-source software library available at \url{https://github.com/Magritte-code/Magritte-torch}. The models in this paper are available upon reasonable request to the authors.



\bibliographystyle{rasti}
\bibliography{Library_bibtex_20250430} 




\appendix
\section{Extrapolation of CO collisional rates}\label{appendix: CO collisional rate extrapolation}
\begin{figure}
	\includegraphics[width=\linewidth]{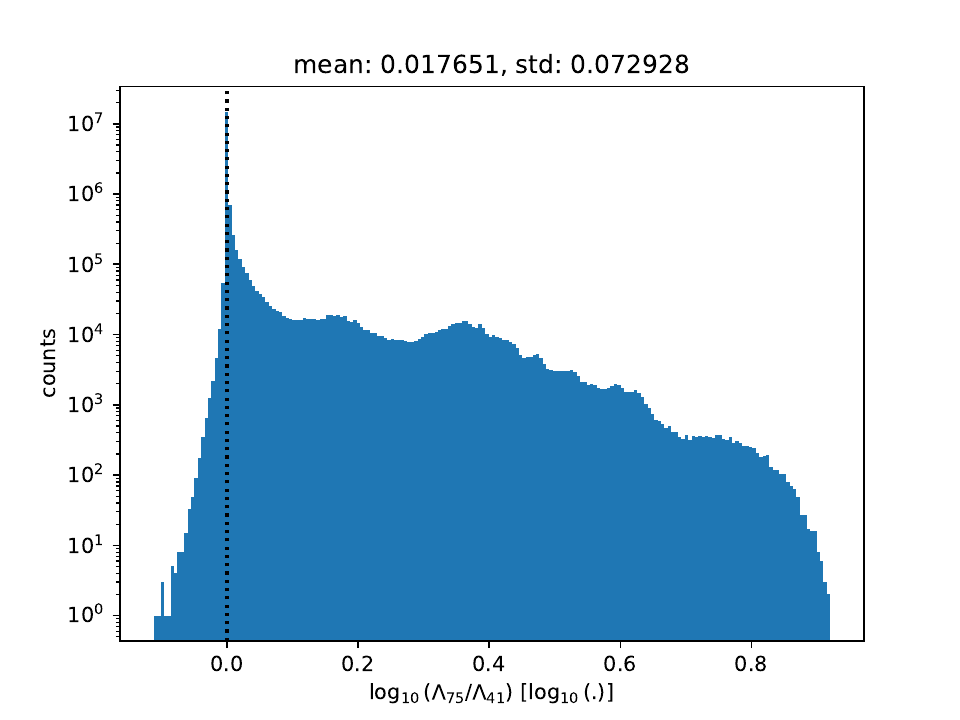}
	\caption{Logarithm of the relative differences between the computed cooling rates, when using the extrapolated CO line data for $75$ rotational levels, compared to the base CO line data, using $41$ rotational levels, as available in the LAMDA database. Note that the y-axis is also in log space to show the outliers.}\label{fig: reldiff 75 vs 41 levels}
\end{figure}

\begin{figure}
	\includegraphics[width=\linewidth]{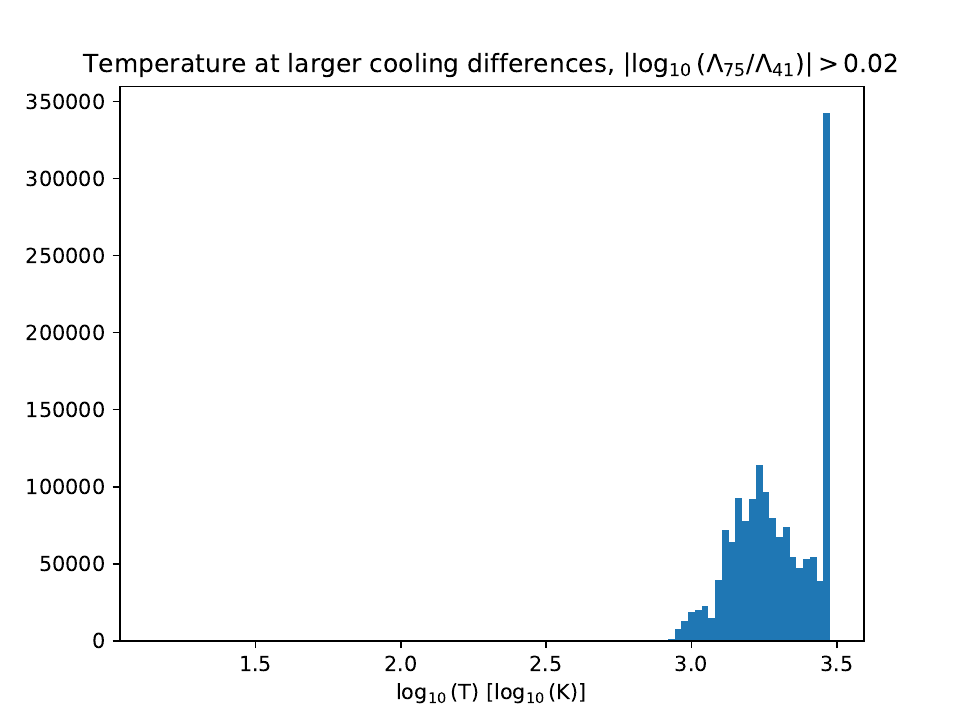}
	\caption{Histogram of the temperatures of data points where the relative differences for the obtained cooling rate $\left|\log_{10}(\Lambda_{75}/\Lambda_{41})\right|$ are significant ($>0.02$) when using the different line data.}\label{fig: large differences 75 vs 41 temperature location}
\end{figure}

The currently available CO line data file with largest available temperature range \citep[using collisional rates based on][]{yang_rotational_2010} includes the first 41 rotational levels, which corresponds to an excitation temperature of about $4500$ K for the highest level. Given that we encounter temperatures of several thousand Kelvin in our hydrodynamics models of AGB stars, the populations of the CO energy levels above the 41st one will be non-negligible.
In this appendix, we clarify how we generate the line data file for CO. For this, we use the energy levels and v=0 line transitions from CDMS, pruned up to 75 levels, which results in an upper level excitation temperature of $15000$ K. For the collisional rates, we start from \cite{yang_rotational_2010}, which gives data for the transitions between the first $41$ rotational levels for temperature between $2$ K and $3000$ K, and extrapolate them using the method described in the appendix of \cite{schoier_atomic_2005}.
 Evidently, we only use this extrapolated data starting from level $42$ to fill in the missing data.
Comparing the 75 level file to the original CO line data, we found a mean relative difference in the computed cooling $\left|\log_{10}(\Lambda_{75}/\Lambda_{41})\right|$ of $0.018$, with standard deviation $0.07$ (see Fig \ref{fig: reldiff 75 vs 41 levels}), in which $\Lambda_{75} = \Lambda_{\text{this work}}$ and $\Lambda_{41}$ are respectively the cooling rates computed using the $75$ and $41$ rotational level CO line data files. Upon further inspection (see Fig. \ref{fig: large differences 75 vs 41 temperature location}), this change practically only impacted the high temperature regions ($>1000K$), which is as we expect.

The extrapolated CO file in LAMDA format \citep[][]{schoier_atomic_2005} and python files used to generate it are available in the additional supplementary data.

\section{Applying the full fit to single star models}\label{appendix: applying full fit to single star}
In the main text, we focused on creating a CO cooling rate fit for hydrodynamics models of AGB stars with a companion. Evidently, AGB stars can also exist without a stellar mass companion object. Therefore, we verify in this appendix whether our fit prescriptions also work for single star models.\\

\begin{figure}
	\includegraphics[width=\linewidth]{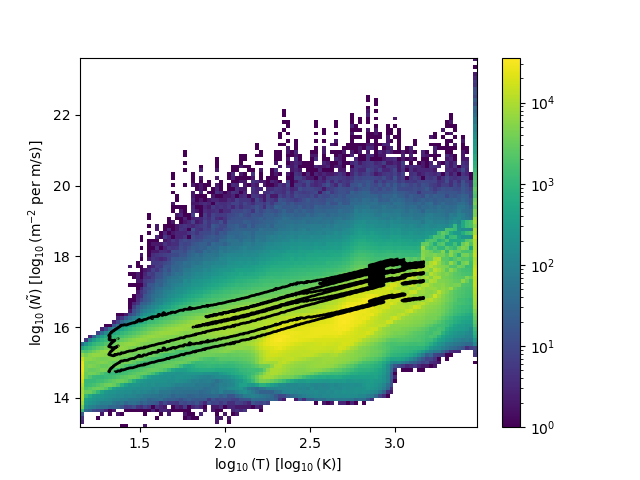}
	\caption{Distribution of our AGB binary model points. The black contours denote the extent of the parameter range for the single star model points.}\label{fig: parameter space single star subset}
\end{figure}

\begin{figure}
	\includegraphics[width=\linewidth]{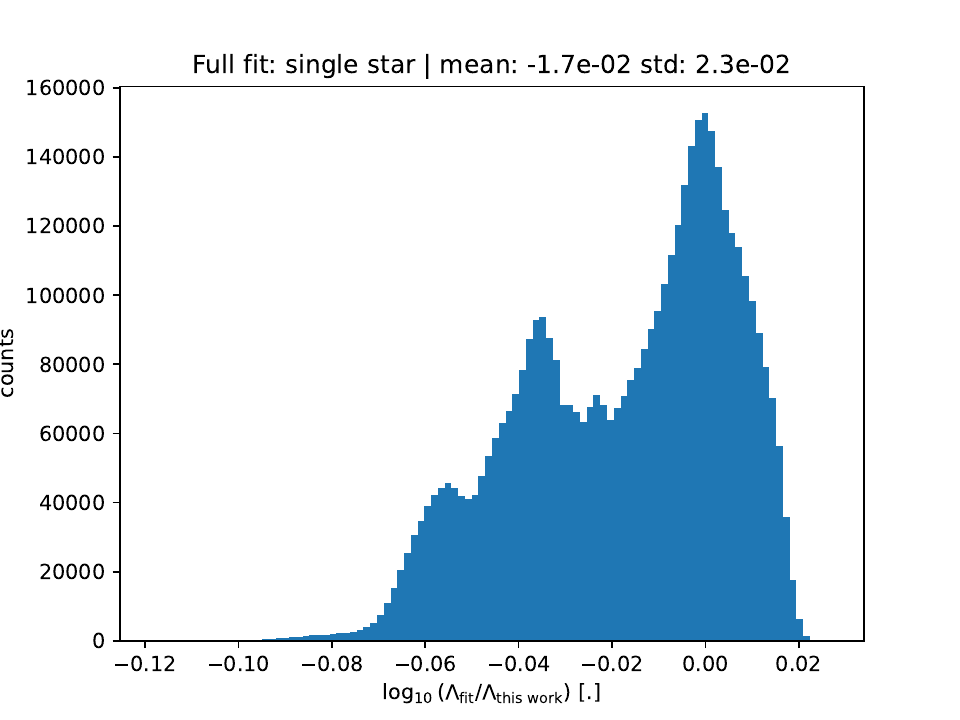}
	\caption{Histogram of relative differences between the reference results for the single star data and our CO cooling cooling fit, in log space.}\label{fig: full fit single star}
\end{figure}

In Section \ref{section: general model setup}, we have explained how we created the radiative transfer models for all AGB models in this paper, including the single star models. Given that the parameter space of data points within the single star model is a subset of the parameter space of data points for the binary models (see Fig. \ref{fig: parameter space single star subset}), we expect that our fits of Section \ref{section: fit all} will be adequate for the single star models. In Fig. \ref{fig: full fit single star}, we confirm this by applying both the low and high density fits (in their relevant parameter ranges) to the single star data, obtaining a mean relative difference for the log cooling rate $\log_{10}\left(\Lambda_{\text{fit}}/\Lambda_{\text{this work}}\right)$ of $-1.7\cdot 10^{-2}$ dex, with standard deviation of $2.3\cdot 10^{-2}$ dex.

\section{Existing cooling prescriptions within and outside their intended parameter range}\label{appendix: existing cooling prescriptions within and outside their parameter range}
\begin{figure}
	\includegraphics[width=\linewidth]{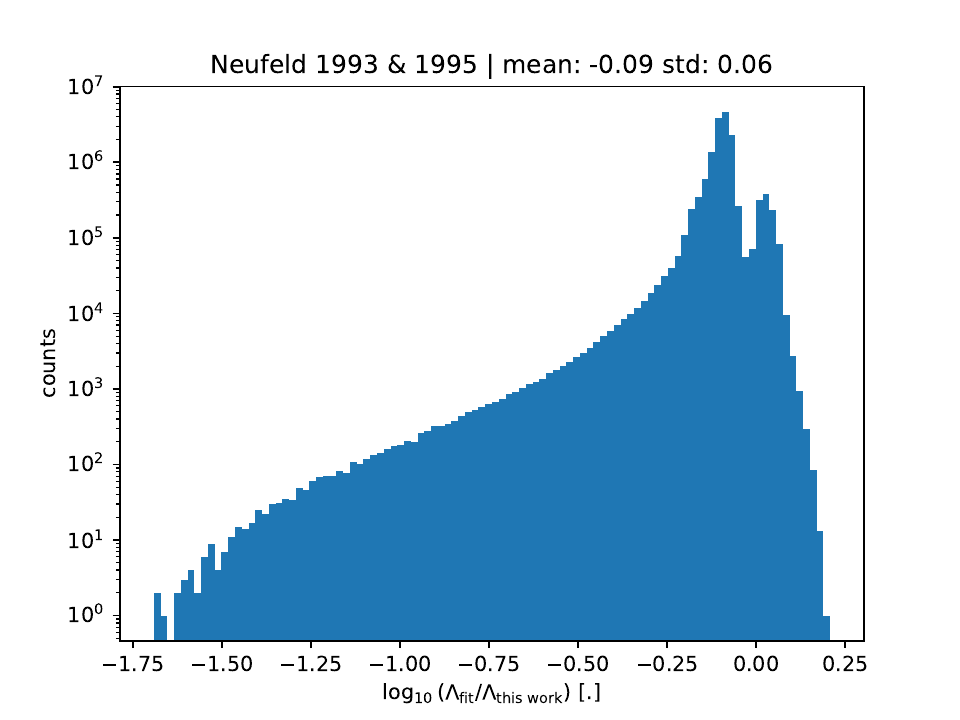}
	\caption{Histogram of relative differences between the reference results and the cooling fit result using the combined Neufeld prescriptions \citep{neufeld_radiative_1993, neufeld_thermal_1995}, in log space. Note that the y-axis is also in log space to show the outliers. Input data to limited to the intended parameter regime.}\label{fig: neufeld within param bounds}
\end{figure}

\begin{figure}
	\includegraphics[width=\linewidth]{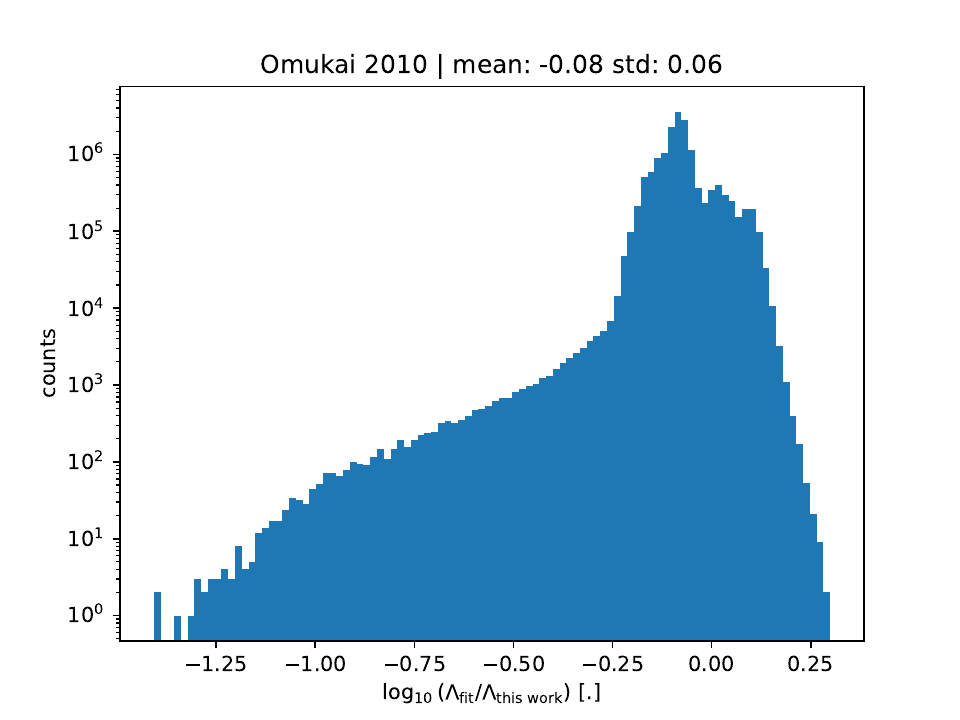}
	\caption{Histogram of relative differences between the reference results and the cooling fit result using the \protect\cite{omukai_low-metallicity_2010} prescription, in log space. Note that the y-axis is also in log space to show the outliers. Input data to limited to the intended parameter regime.}\label{fig: omukai within param bounds}
\end{figure}

\begin{figure}
	\includegraphics[width=\linewidth]{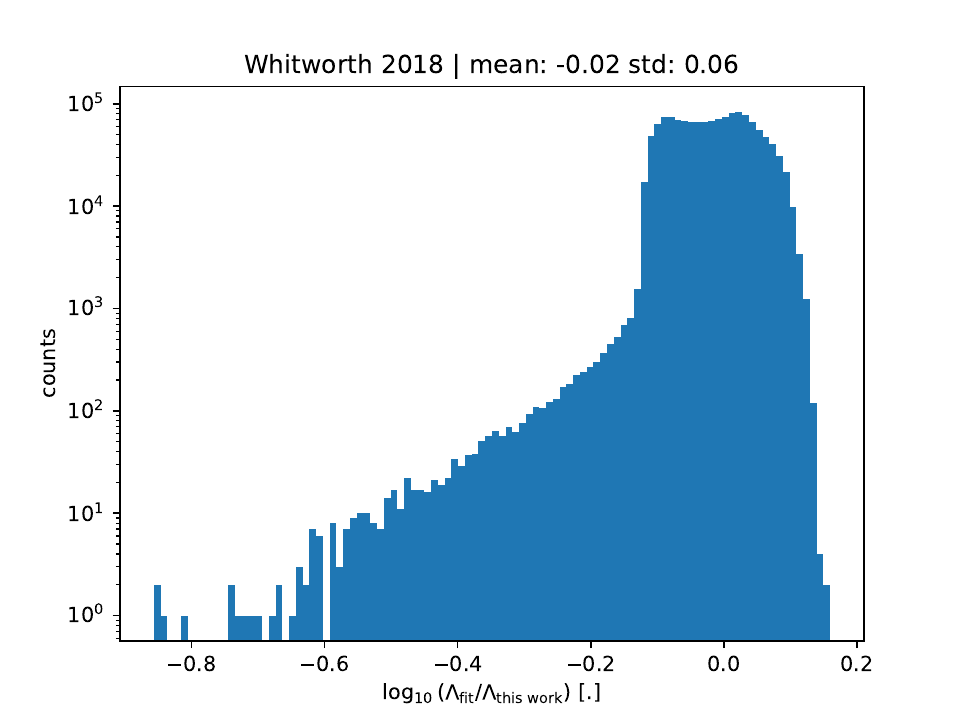}
	\caption{Histogram of relative differences between the reference results and the cooling fit result using the \protect\cite{whitworth_simple_2018} prescription, in log space. Note that the y-axis is also in log space to show the outliers. Input data to limited to the intended parameter regime.}\label{fig: whitworth within param bounds}
\end{figure}

\begin{figure}
	\includegraphics[width=\linewidth]{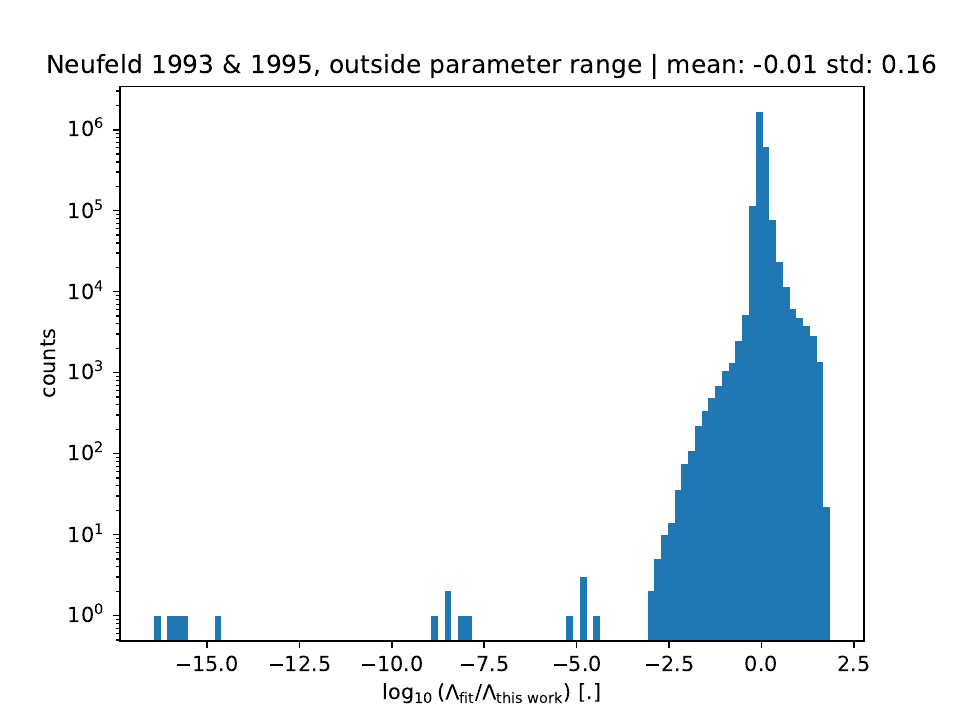}
	\caption{Histogram of relative differences between the reference results and the cooling fit result using the combined Neufeld prescriptions \citep{neufeld_radiative_1993, neufeld_thermal_1995}, in log space. Note that the y-axis is also in log space to show the outliers. Input data to limited to \textbf{outside} the intended parameter regime.}\label{fig: neufeld outside param bounds}
\end{figure}

\begin{figure}
	\includegraphics[width=\linewidth]{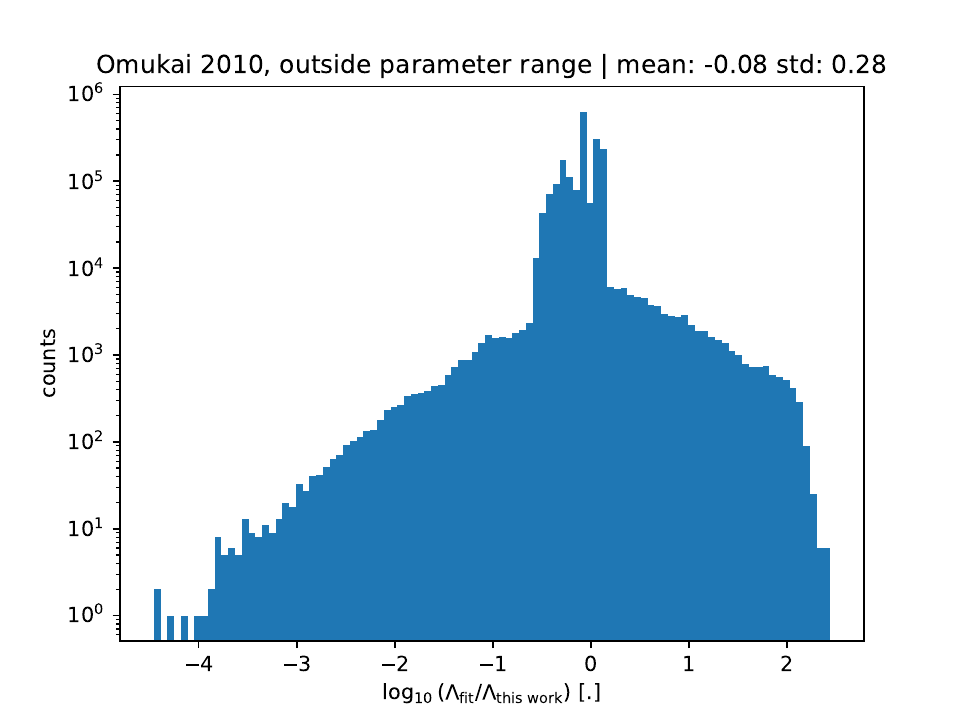}
	\caption{Histogram of relative differences between the reference results and the cooling fit result using the \protect\cite{omukai_low-metallicity_2010} prescription, in log space. Note that the y-axis is also in log space to show the outliers. Input data to limited to \textbf{outside} the intended parameter regime.}\label{fig: omukai outside param bounds}
\end{figure}

\begin{figure}
	\includegraphics[width=\linewidth]{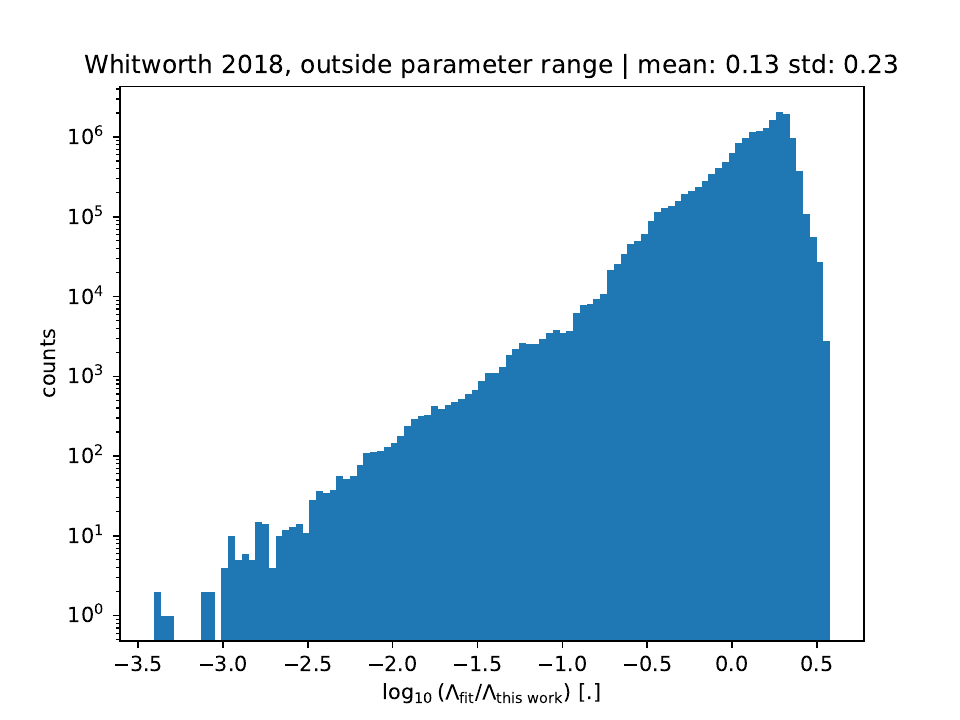}
	\caption{Histogram of relative differences between the reference results and the cooling fit result using the \protect\cite{whitworth_simple_2018} prescription, in log space. Note that the y-axis is also in log space to show the outliers. Input data to limited to \textbf{outside} the intended parameter regime.}\label{fig: whitworth outside param bounds}
\end{figure}

In the main text, we argued that the existing CO cooling prescriptions are not fully applicable to the AGB environment, given that the intended parameter regimes are smaller than the parameter space we consider. Thus, for completeness, we add here the results applied to respectively our data points within and outside the originally intended parameter regimes. Figs. \ref{fig: neufeld within param bounds}, \ref{fig: omukai within param bounds}, and \ref{fig: whitworth within param bounds} show that these prescriptions work well inside their regime, while Figs. \ref{fig: neufeld outside param bounds} and \ref{fig: omukai outside param bounds} show that the combined Neufeld \citep{neufeld_radiative_1993, neufeld_thermal_1995} and \cite{omukai_low-metallicity_2010} prescriptions perform remarkably worse outside this regime. Given that given that most of our data points ($\sim 92\%$) lie outside the parameter range of the \cite{whitworth_simple_2018} prescription (See Fig. \ref{figure: parameter_bounds_whitworth}), not much can be inferred from Fig. \ref{fig: whitworth outside param bounds}, apart from the fact that the prescription still seems to perform reasonably, even so far outside the intended parameter bounds.

\bsp	
\label{lastpage}
\end{document}